\documentclass[a4paper,oneside,11pt]{amsart}

\usepackage{amsmath,amssymb,epsf}
\usepackage{amsfonts,amsbsy,amscd,stmaryrd}
\usepackage{latexsym,amsopn}
\usepackage{amsthm}

\usepackage{graphicx}
\usepackage{color}
\definecolor{Blue}{rgb}{0.,0.,1.}
\definecolor{Red}{rgb}{1.,0.,0.}

\newcounter{smallarabics}
\newenvironment{arabicenumerate}
{\begin{list}{{\normalfont\textrm{(\arabic{smallarabics})}}}
  {\usecounter{smallarabics}\setlength{\itemindent}{0cm}
   \setlength{\leftmargin}{5ex}\setlength{\labelwidth}{4ex}
   \setlength{\topsep}{0.75\parsep}\setlength{\partopsep}{0ex}
   \setlength{\itemsep}{0ex}}}
{\end{list}}

\newcounter{smallroman}

\newcommand{\ben}{\begin{arabicenumerate}}  
\newcommand{\een}{\end{arabicenumerate}}

\def\init{\setcounter{equation}{0}}

\newtheorem{theoreme}{Theorem}[section]
\newtheorem{proposition}[theoreme]{Proposition}
\newtheorem{assumption}[theoreme]{Hypothesis}
\newtheorem{lemma}[theoreme]{Lemma}
\newtheorem{definition}[theoreme]{Definition}

\newtheorem{remark}[theoreme]{Remark}
\newtheorem{example}[theoreme]{Example}
\newcommand{\beq}{\begin{equation}}
\newcommand{\eeq}{\end{equation}}

\newcommand{\bex}{\begin{example}}
\newcommand{\eex}{\end{example}}
\def\bel{\begin{lemma}}
\def\eel{\end{lemma}}
\def\bet{\begin{theoreme}}
\def\eet{\end{theoreme}}
\def\bed{\begin{definition}}
\def\eed{\end{definition}}
\def\ber{\begin{remark}}
\def\eer{\end{remark}}

\def\rr{{\mathbb R}}
\def\zz{{\mathbb Z}}
\def\cc{{\mathbb C}}

\def\part{{\rm par}}

\def\bar{\overline}

\def\c0inf{C_0^\infty}

\def\proof{
\noindent{\bf Proof.}\ \ }

\usepackage{calrsfs}
\DeclareMathAlphabet{\pazocal}{OMS}{zplm}{m}{n}

\def\cV{{\pazocal V}}

\def\kL{{\mathcal L}}

\def\cN{{\pazocal N}}

\def\fg{{\mathfrak g}}

\def\CCR{{\rm CCR}}
\def\CAR{{\rm CAR}}
\def\aCCR{{\mathfrak{A}_{\CCR}^{\rm pol}}}
\def\aCAR{{\mathfrak{A}_{\CAR}^{\rm pol}}}
\def\aCCAR{{\mathfrak{A}_{\CCR/\CAR}^{\rm pol}}}
\def\aGH{{\mathfrak{A}_{\gh}^{\rm pol}}}
\def\wf{{\rm WF}}

\def\i{{\rm i}}
\def\ad{{\rm ad}}

\DeclareMathOperator{\Ker}{Ker}
\DeclareMathOperator{\Ran}{Ran}

\renewcommand{\c}{{\rm c}}
\newcommand{\qeds}{\qed\medskip}

\def\Int{\rm Int}

\def\12{\frac{1}{2}}
\def\14{\frac{1}{4}}

\def\ad{{\rm ad}}

\newcommand{\one}{\boldsymbol{1}}
\def\cH{{\pazocal H}}

\def\cK{{\pazocal K}}

\def\Int{\rm Int}

\def\12{\frac{1}{2}}

\def\ad{{\rm ad}}

\def\Diff{{\rm Diff}}

\def\bep{\begin{proposition}}
\def\eep{\end{proposition}}

\def\gh{{\scriptstyle\#}{\rm gh}}

\def\CARal{{\rm C\hskip 0.25 em \hbox{\raise 1.72 ex 
\hbox{$\scriptscriptstyle\rm al$}\kern -0.57 em A}R}}

\def\otimesal{\mathop{\hbox{\raise 1.5 ex
  \hbox{$\scriptscriptstyle\rm al$}
\kern -0.92 em \hbox{$\otimes$}}}}
\def\oplusal{\mathop{\hbox{\raise 1.5 ex
  \hbox{$\scriptscriptstyle\rm al$}
\kern -0.92 em \hbox{$\oplus$}}}}
\def\Gammal{\hbox{\raise 1.68 ex 
\hbox{$\scriptscriptstyle\rm al$}\kern -0.50 em $\Gamma$}}
\def\Bal{\hbox{\raise 1.68 ex 
\hbox{$\scriptscriptstyle\rm  al$}\kern -0.50 em $B$}}
\def\CARal{{\rm C\hskip 0.25 em \hbox{\raise 1.72 ex 
\hbox{$\scriptscriptstyle\rm al$}\kern -0.57 em A}R}}

\renewcommand{\Int}{\, \lrcorner \,}
\newcommand{\tnI}{\, \llcorner \,}

\DeclareMathAlphabet{\mathpzc}{OT1}{pzc}{m}{it}

\DeclareSymbolFont{boldoperators}{OT1}{cmr}{bx}{n}
\SetSymbolFont{boldoperators}{bold}{OT1}{cmr}{bx}{n}
\edef\wbar{\unexpanded{\protect\mathaccentV{bar}}\number\symboldoperators16}

\def\bds{\wbar{d}_{\Sig}}

\def\bdeltas{{\wbar{\delta}_{\Sig}}}

\def\ud{d}

\def\bds{\wbar{d}_{\Sig}}

\def\bdeltas{{\wbar{\delta}_{\Sig}}}

\def\Gsc{{\Gamma_{\rm sc}}}
\def\Gc{\Gamma_{\rm c}}

\def\pback{{\scriptscriptstyle *}}

\newcommand*{\remthinspace}{\hskip-0.05em\relax}
\def\Sig{{\remthinspace\scriptscriptstyle\Sigma}}
\def\sD{{\remthinspace\scriptscriptstyle D}}
\def\sL{{\remthinspace\scriptscriptstyle L}}
\def\sA{a}
\def\sB{b}
\def\sDo{{\remthinspace\scriptscriptstyle D}}
\def\sP{{\remthinspace\scriptscriptstyle P}}
\def\sQ{{\remthinspace\scriptscriptstyle Q}}
\def\sR{{\remthinspace\scriptscriptstyle R}}
\def\sV{{\scriptscriptstyle V}}
\def\sVA{{\scriptscriptstyle V}}
\makeatletter
\newcommand*{\defeq}{\mathrel{\rlap{%
                     \raisebox{0.3ex}{$\m@th\cdot$}}%
                     \raisebox{-0.3ex}{$\m@th\cdot$}}%
                     =}
\makeatother
\makeatletter
\newcommand*{\eqdef}{=\mathrel{\rlap{%
                     \raisebox{0.3ex}{$\m@th\cdot$}}%
                     \raisebox{-0.3ex}{$\m@th\cdot$}}%
                     }
\makeatother

\DeclareMathAlphabet{\mathpzc}{OT1}{pzc}{m}{it}
\def\killing{\mathpzc{k}}
\def\gi{{\rm (g.i.)}}
\def\pos{{\rm (pos)}}
\def\musc{(\mu{\rm sc})}

\def\gho{{|_{[0]}}}
\def\Gho{{\Big|_{[0]}}}

\def\Ghi{{\Big|_{[i]}}}

\newcommand{\rhs}{r.h.s.\ }

\newcommand{\wrt}{w.r.t.\ }

\newcommand{\scalP}[2]{\left( #1 , #2 \right)}

\renewcommand{\slash}[2][4]{\ensuremath{\rlap{\raisebox{1pt}{$\mspace{#1mu}/$}}#2}}

\newcommand{\nabslash}{\slash[3]{\nabla}}

\newcommand{\DM}{{\rm\textit{DM}}}
\newcommand{\RS}{{\scriptscriptstyle\rm RS}}
\newcommand{\HaS}{{\scriptscriptstyle\rm HS}}

\begin{document}
\title[BRST formalism for gauge field theories on curved spacetime]{{\large Classical phase space and Hadamard states} \\  {\large in the BRST formalism for gauge field theories} \\  {\large  on curved spacetime}}
\author{}
\address{Universit\'e Grenoble Alpes, CNRS, Institut Fourier, F-38000 Grenoble, France}
\email{michal.wrochna@univ-grenoble-alpes.fr}
\author{\normalsize Micha{\l} \textsc{Wrochna}  \& Jochen \textsc{Zahn}}
\address{Universit\"at Leipzig, Institut f\"ur Theoretische Physik,  Br\"uderstr. 16, 04103 Leipzig, Germany}
\email{jochen.zahn@itp.uni-leipzig.de}

\begin{abstract}
We investigate linearized gauge theories on globally hyperbolic spacetimes in the BRST formalism. A consistent definition of the classical phase space and of its Cauchy surface analogue is proposed.
We prove that it is isomorphic to the phase space in the `subsidiary condition' approach of Hack and Schenkel in the case of Maxwell, Yang-Mills, and Rarita-Schwinger fields. Defining Hadamard states in the BRST formalism in a standard way, their existence in the Maxwell and Yang-Mills case is concluded from known results in the subsidiary condition (or Gupta-Bleuler) formalism. Within our framework, we also formulate criteria for non-degeneracy of the phase space in terms of BRST cohomology and discuss special cases. These include an example in the Yang-Mills case, where degeneracy is not related to a non-trivial topology of the Cauchy surface. 
\end{abstract}

\let\origmaketitle\maketitle
\def\maketitle{
  \begingroup
  \def\uppercasenonmath##1{} 
  \let\MakeUppercase\relax 
	\origmaketitle
  \endgroup
}

\maketitle

\section{Introduction \& summary}\init\label{sec:intro}

\subsection{Introduction} The \emph{Becchi-Rouet-Stora-Tyutin} or in short, the \emph{BRST formalism} \cite{BRS,tyutin},  is nowadays regarded as an essential ingredient in the perturbative quantization of gauge field theories. The algebraic structures it relies on have been extensively studied in the literature \cite{BBH,HT,HT2} and their incorporation in perturbative interacting theories on curved spacetime has been achieved by Hollands in \cite{hollands}, followed by recent works on the more general Batalin-Vilkovisky formalism \cite{FR,rejzner}.

The basis of the perturbative quantization is the linearized theory, and in the present work we investigate its kinematical content and demonstrate that the BRST quantization can be formulated using the standard apparatus of algebraic quantum field theory on curved spacetime, i.e. via states on a $*$-algebra (or $C^*$-algebra) of canonical (anti)-commutation relations. Thus, the classical non-interacting theory is described by a phase space $(\cV,q)$ (a vector space $\cV$ equipped with a hermitian form $q$), and the physical Hilbert space is obtained by GNS construction after choosing a state $\omega$ on the CCR or CAR $*$-algebra associated to $(\cV,q)$. Moreover, a conventional definition of \emph{Hadamard states} ensures that they enjoy the properties needed to construct the perturbative interacting theory. 

In contrast to the existing literature, we do not start from a Lagrangean formulation. Instead we just assume that the equations of motion are given by a differential operator $L$, that already contains the unphysical degrees of freedom,  together with another differential operator $\gamma$ which generates the BRST symmetry. Assuming the pair $L,\gamma$ satisfies a number of conditions (typically fulfilled in any linear system coming from a BRST Lagrangean), we construct the classical phase space $(\cV,q)$ and find spaces isomorphic to it, expressed in terms of space-compact solutions of $L$ and their Cauchy data. 


\subsection*{Overview of BRST formalism} To motivate our framework and the properties of $L,\gamma$ assumed in the main part of the text, let us recall the basic ingredients of the BRST formalism. 

Let $(M,g)$ be a globally hyperbolic space-time, $V_1$ a finite-rank bundle over $M$ with hermitian structure $(\cdot|\cdot)_{V_1}$. Suppose we are given a Lagrangian $\kL$, in general non-linear, whose variation gives the equations of motion operator
\[
P: \Gamma(M;V_1) \to \Gamma(M;V_1),
\]
acting on smooth sections $\Gamma(M;V_1)$ of $V_1$.
Furthermore, one assumes there is a gauge symmetry, i.e., a group $G$ acting on $\Gamma(M;V_1)$ such that
\[
 P(\varphi) = 0 \ \Leftrightarrow \ P(g \varphi) = 0 \quad \forall\,g\in G.
\]
The corresponding Lie algebra is assumed to be isomorphic to $\Gamma(M; V_0)$ for some bundle $V_0$. Hence, there is a local operator
\[
 K: \Gamma(M;V_1) \times \Gamma(M;V_0) \to \Gamma(M;V_1),
\]
which is linear in its second argument, and fulfills
\begin{equation*}
 P'(\varphi)(K(\varphi, f)) = 0 \quad \forall\,\varphi \in \Gamma(M;V_1) \ {\rm s.t.} \ P(\varphi) = 0, \,f \in \Gamma(M; V_0),
\end{equation*}
where $P'(\varphi)$ is the differential of $P$ at $\varphi \in \Gamma(M; V_1)$. Taking $f$ to be compactly supported, one concludes that the linearized wave operator $P'(\varphi)$ is not hyperbolic (which is the main difficulty).

The BRST formalism amounts to introducing auxiliary degrees of freedom, termed Lagrange multipliers $b$, ghosts $c$, and antighosts $\bar c$. This means that one considers an enlarged vector bundle $V$, obtained by taking the direct sum of $V_1$ and (typically) $V_0^{\oplus 3}$.  In addition, to keep track of different types of degrees of freedom, one introduces a grading $\gh$ called the \emph{ghost number}. Conventionally, physical degrees of freedom and Lagrange multipliers $b$ correspond to ghost number $0$, whereas ghosts $c$ (anti-ghosts $\bar{c}$) have ghost number $1$ ($-1$).

One also introduces supplementary terms to the Lagrangian. Assume that $P(\varphi) = 0$ has, at least locally, a well posed Cauchy problem given the gauge fixing condition $T(\varphi) = 0$, where $T: \Gamma(M;V_1) \to \Gamma(M; V_0)$. Then define a new Lagrangean
\[
 \kL_{\rm\scriptscriptstyle BRST}(f) \defeq \kL(\varphi) + ( f_{\bar c} |  T(K(\varphi, f_c)))_{V_0} + (f_b| T(\varphi))_{V_0} + \tfrac{\alpha}{2} (f_b| R f_b)_{V_0}, 
\]
where $R$ is some suitable differential operator and $f=(\varphi,f_b,f_c,f_{\bar{c}})\in\Gamma(M;V)$. Typically, the  linearized equations of motion for such choice of Lagrangean are given by a differential operator, denoted $L\in\Diff(M;V)$, which is \textit{hyperbolic} and \textit{preserves the grading}. 

Next, one introduces the \textit{BRST operator}\footnote{In our convention $\gamma$ is the formal adjoint (or transpose) of the BRST differential used in most of the literature, note that it acts on configurations instead of evaluation functionals.} $\gamma$, which in the setting linearized around a solution $\varphi$ acts by
\[
\gamma (f_a,f_b,f_c,f_{\bar{c}}) = (K(\varphi, f_c), 0, 0, f_b)
\]
(where $f_a$ is the linear perturbation of $\varphi$). It is a nilpotent symmetry of $L$ in the sense that
\beq\label{eq:ns}
\gamma^2=0, \quad \gamma^* L = L \gamma,
\eeq
and it \emph{decreases the grading by one}. Formally, the physical degrees of freedom are recovered by restricting to solutions of $L$ with ghost number $0$  and then taking the quotient space $\Ker \gamma/\Ran \gamma$. In this paper we argue that the correct choice of physical phase space is given rigorously by the restriction of
\beq\label{eq:phsp1}
\frac{\Ker L|_{\Gsc}\cap \Ker \gamma|_{\Gsc}}{\Ran G_\sL\gamma^*|_{\Gc}}
\eeq
to ghost number 0 sections, where $G_\sL$ is the causal propagator (Pauli-Jordan commutator function) of $L$, and the notation $\Gamma_{\rm c}$, $\Gamma_{\rm sc}$ refers to compactly supported, resp. space-compact smooth sections.
First, we prove that the causal propagator $G_\sL$ induces a well-defined (anti-)hermitian form on the above quotient for any pair $L,\gamma$ satisfying \eqref{eq:ns} and a few further properties.
Moreover, we show that the so-defined phase space is isomorphic to the ghost number 0 restriction of $\Ker \gamma_\Sig/\Ran \gamma_\Sig$ for some operator $\gamma_\Sig$ acting on smooth, compactly supported Cauchy data of $L$. 

Our presentation of the subject is focused on the ingredients of the BRST formalism that are needed to construct Hadamard states, for instance the Cauchy surface version of the phase space is essential to use the methods of \cite{GW2}. 





\subsection*{Relation to other frameworks} Most of the existing literature on gauge theories on curved spacetime uses various versions of an approach called in this work the \emph{subsidiary condition framework}\footnote{Its essential feature is that the kinematics are given by a hyperbolic PDE with a constraint, often called `subsidiary condition' in the literature.} \cite{dimock2,SDH,DS,FP,GW2,igorpreprint,igorgauge,HS,pfenning,FS}, or the very closely related \emph{Gupta-Bleuler formalism} \cite{FS}. A general formulation has been recently proposed by Hack and Schenkel \cite{HS} and one of our goals is to {relate} it with the BRST formalism. As anticipated \cite{hollands}, in the case of the Maxwell and Yang-Mills equation in the Feynman gauge there is a direct relation, both on the level of phase spaces and states. It turns out that such kind of relation can also be derived for the Rarita-Schwinger equation. 

\subsection*{Degeneracy of the phase space} An issue that has recently attracted wide interest is the possible degeneracy of the phase space $(\cV,q)$ if the Cauchy surface $\Sigma$ is topologically non-trivial \cite{SDH,HS,benini,igorgauge,khavkine}. Specifically, it is known in the Maxwell case that $q$ is non-degenerate on $\cV$ if and only if
\beq\label{eq:knownmaxwell}
\Ran d_\Sig^0|_{\Gamma_{\rm c}} = \Ran d_\Sig^0|_{\Gamma} \cap \Gamma_{\rm c}(\Sigma;\Lambda^1),
\eeq
where $d_\Sig^0|_{\Gamma}$ (resp. $d_\Sig^0|_{\Gamma_\c}$) is the differential acting on smooth (resp. smooth, compactly supported) $0$-forms \cite{SDH}. We show that in the BRST framework an analogous result in terms of $\gamma_\Sig$ (strictly speaking its formal adjoint $\gamma_\Sig^*$) holds true assuming a \emph{generalized Poincar\'e duality}. The key observation is that properties such as (\ref{eq:knownmaxwell}) amount to injectivity of canonical maps between $d_\Sig$-cohomology of different types: compactly supported cohomology, de Rham (i.e. smooth), distributional, etc. In the BRST formalism it is possible to use $\gamma_\Sig^*$-cohomology instead.  

 
From considerations on compactly supported $\gamma_\Sig^*$-cohomology it turns out that the Yang-Mills equation linearized around an on-shell non-trivial background connection reveals new features, not present for flat background connections: we find specifically that degeneracy of $q$ is well possible even if $\Sigma$ is topologically trivial.

\subsection{Outlook} In the present work we study among other the issue of degeneracy of the phase space by means of BRST cohomology. An open question necessary to derive more explicit results is the validity of the generalized Poincar\'e duality introduced in Subsect. \ref{ss:nond}, for theories such as the Yang-Mills equation linearized around a non-trivial solution. One difficulty appears to be the non-ellipticity of the complex associated to $\gamma_\Sig$, in the sense that distributional are not naturally identified with a space of smooth sections (in contrast to de Rham theory). It is therefore possible that the existence of an appropriate elliptic complex could be helpful, as for instance the twisted de Rham complex proposed in \cite{khavkine,igorcalabi} for the Yang-Mills equation.

The main purpose of our paper is to provide the basic ingredients needed to construct Hadamard states in the BRST framework. A particularly interesting problem that remains open is the existence of Hadamard states for the Rarita-Schwinger equation, although in view of our results it is sufficient to derive a construction in the subsidiary condition framework.

Important examples of gauge theories not discussed in the present paper include linearized gravity, cf. the recent works \cite{FH,khavkine,BDM}, and the perturbative quantization of the Nambu-Goto string as formulated in \cite{BRZ}, we expect however that our results apply as well. The rigorous construction of states in the BRST formalism can lead to interesting issues, especially in view of the difficulties found for linearized gravity in the subsidiary condition framework in \cite{BDM}. 

\subsection{Plan of the paper} The paper is structured as follows.

Sect.~\ref{sec:classical} is focused on classical gauge field theories. After recalling some preliminaries, we review in Subsect.~\ref{ss:subsidiary} the subsidiary condition framework of Hack and Schenkel. Subsect.~\ref{ss:BRST} is the key part of paper, in which we introduce our abstract version of the BRST formalism, and derive equivalent formulae for the physical phase space. The issue of its (non)-degeneracy is discussed in Subsect.~\ref{ss:nond}. We prove therein a criterion (Thm.~\ref{thm:criterion}) in terms of compactly supported $\gamma_\Sig^*$-cohomology. Next, we show in Subsect.~\ref{ss:relation} how the two frameworks are related. We assume therein a simplified version of Hack and Schenkel's framework, which includes the case of Maxwell and Yang-Mills fields.

In Sect. \ref{sec:states} we define Hadamard states in a standard way, and show that the relation between the two frameworks extends to the level of states.

Sect. \ref{s:examples} gathers examples of applications of our framework. In Subsect. \ref{ss:max} we focus on the Maxwell field, and show that in that case our criterion for non-degeneracy of the symplectic form (Thm. \ref{thm:criterion}) reduces to conditions on the usual compactly supported and de Rham cohomology. In Subsect. \ref{ss:YM} we consider the Yang-Mills equation linearized around a non-trivial solution and show an example of degeneracy of the phase space. We then discuss in Subsect.~\ref{ss:RS} the Rarita-Schwinger equation in the BRST and subsidiary condition framework.  

\section{Two formalisms for classical gauge field theories}\label{sec:classical}

\subsection{Notations --- differential operators} 
Let $V,W$ be vector bundles over a smooth manifold\footnote{We always consider complex vector bundles of finite rank, smooth manifolds are always assumed to be Hausdorff.} $M$. Smooth sections of $V$ will be denoted $\Gamma(M;V)$, and compactly supported ones ${\Gc}(M;V)$. The set of differential operators (of order $m$) $\Gamma(M;V)\to\Gamma(M;W)$ is denoted  $\Diff(M;V,W)$ ($\Diff^m(M;V,W)$), we also set $\Diff(M;V)=\Diff(M;V,V)$. 

By a \emph{bundle with hermitian structure} we will mean a vector bundle $V$ equipped with a fiberwise non-degenerate hermitian form $(\cdot, \cdot)_V$ (we do not assume it is positive definite). 

Suppose that $(M,g)$ is a pseudo-Riemannian oriented manifold. If $V$ is a vector bundle on $M$ with hermitian structure, we denote $V^*$ the anti-dual bundle. The hermitian structure on $V$ and the volume form on $M$ allow to embed $\Gamma(M; V)$ into $\Gamma_{\rm c}'(M; V)$, using the non-degenerate hermitian form on $\Gamma_{\rm c}(M; V)$
\begin{equation}
\label{defdepscalM}
(u|v)_{V}\defeq  \int_{M}(u(x), v(x))_{V}d{\rm Vol}_{g}, \ u, v\in \Gamma_{\rm c}(M; V).
\end{equation}
induced from the hermitian form $(\cdot,\cdot)_{V}$ on fibers. The formal adjoint of an operator $A:{\Gc}(M;V)\to\Gamma(M;W)$ with respect to $(\cdot|\cdot)_V$ is denoted $A^*:{\Gc}(M;W)\to\Gamma(M;V)$.

If $E,F$ are vector spaces, the space of linear operators is denoted $L(E,F)$. If $E,F$ are additionally endowed with some topology, we write $A:E\to F$ if $A\in L(E,F)$ is continuous.

To distinguish between the same operator $A$ acting on different spaces of functions and distributions, for instance $A:\Gamma_{\rm c}(M;V)\to\Gamma_{\rm c}'(M;W)$ and $A:\Gamma(M;V)\to\Gamma(M;W)$, we use the notation $A|_{\Gamma_{\rm c}}$ and $A|_{\Gamma}$. We stress that accordingly, $\Ran A|_{\Gc}$ is in general not the same space as $(\Ran A|_{\Gamma})\cap\Gamma_{\rm c}$. 

\subsection{Quotient spaces}

In the sequel we will frequently encounter operators and sesquilinear forms on quotients of linear spaces, we recall thus the relevant basic facts.

\subsubsection{Operators on quotient spaces}

Let $F_i\subset E_i$, $i=1,2$ be vector spaces and let $A\in L(E_1, E_2)$. Then the induced map
\[
[A]\in L( E_1/F_1, E_2/F_2),
\]
defined in the usual way, is
\begin{itemize}
\item well-defined if $A E_1\subset E_2$ and $A F_1\subset F_2$;
\item injective iff $A^{-1}F_2=F_1$;
\item surjective iff $E_2=A E_1+F_2$. 
\end{itemize} 

\subsubsection{Sesquilinear forms on quotients} Let now $E\subset F$ be vector spaces and let $C$ be a sesquilinear form on $E$. Then the induced sesquilinear form $[C]$ on $E/F$ is
\begin{itemize}
\item well-defined if $CE\subset F^{\circ}$ (where $F^\circ$ denotes the annihilator of $F$) and $F\subset\Ker  C$;
\item non-degenerate iff additionally $F=\Ker C$. 
\end{itemize}

If $C$ is hermitian or anti-hermitian (which will often be the case in our examples) then the condition $F\subset\Ker  C$ implies the other one $CE\subset F^{\circ}$ (and vice-versa).

\subsection{Ordinary classical field theory}\label{ssec:classical}
Let $(M,g)$ be a globally hyperbolic spacetime (we use the convention $(-,+,\dots,+)$ for the Lorentzian signature). If $V$ is a vector bundle over $M$, we denote $\Gsc(M;V)$ the space of space-compact sections, i.e. sections in $\Gamma(M;V)$ such that their restriction to a Cauchy surface has compact support.


One says that $D\in\Diff(M;V)$ is \emph{Green hyperbolic} if $D$ and $D^*$ possess retarded and advanced propagators
--- the ones for $D$ will be denoted respectively $G^{+}_\sDo$ and $G^-_\sDo$ (for the definition, see textbooks \cite{BGP,derger}). As shown in \cite{Baer15}, these are unique. The \emph{causal propagator} (or Pauli-Jordan commutator function) of $D$ is by definition $G_\sDo\defeq G^+_\sDo-G^-_\sDo$. 






Before discussing gauge theories, let us recall the basic data that define an ordinary classical field theory (i.e., with no gauge freedom built in) on a globally hyperbolic manifold $(M,g)$.

\begin{assumption}\label{as:ord}Suppose that we are given:
\begin{enumerate}
\item a bundle $V$ over $M$ with hermitian structure;
\item a Green hyperbolic operator $D\in\Diff(M;V)$ s.t. $D^*=D$.
\end{enumerate}
\end{assumption}

The next two propositions are well-known results, see e.g. \cite{BGP}.

\begin{proposition}\label{prop:prebasic}Let $D\in\Diff(M;V)$ be Green hyperbolic. Then
\[
\Ker D|_\Gsc = \Ran G_\sD|_{{\Gc}}.
\]
\end{proposition}

\begin{proposition}\label{prop:basic}Assume Hypothesis \ref{as:ord}, then
\begin{enumerate}
\item\label{prop:basic2} the induced map 
\[
[G_\sDo]:\,\frac{{\Gc}(M;V)}{\Ran D|_{{\Gc}}}\longrightarrow\Ker D|_{\Gsc}
\]
is well defined and bijective.
\item $(G^{\pm}_{\sDo})^*=G^{\mp}_{\sDo}$ and consequently $G^*_{\sDo}=-G_{\sDo}$;
\end{enumerate}
\end{proposition}

By a \emph{phase space} we mean a pair $(\cV,q)$ consisting of a complex vector space $\cV$ and a sesquilinear form $q$ on $\cV$. Actual physical meaning can be associated to $(\cV,q)$ if $q$ is hermitian (and additionally positive if $(\cV,q)$ is meant to describe a fermionic system). Note that in contrast to most of the literature we consider complex vector spaces, which is slightly more convenient in the discussion of states later on. 

The classical phase space associated to $D$ is $(\cV_{\scriptscriptstyle D},q_{\scriptscriptstyle D})$, where
\[
\cV_{\scriptscriptstyle D} \defeq \frac{{\Gc}(M;V)}{\Ran D|_{{\Gc}}}, \quad \bar u \,q_{\scriptscriptstyle D} v\defeq \i (u| [G_\sDo] v)_{V}
\]
By (2) of Prop. \ref{prop:basic} the sesquilinear form $q_{\scriptscriptstyle D}$ is hermitian and it is not difficult to show that it is non-degenerate. 

For further reference, note the following easy lemma that generalizes a result of Dimock \cite{dimock}, see for instance \cite{wrothesis} for the complete proof.

\begin{lemma}\label{lem:prenormally}If $D,\widetilde{D}\in\Diff(M;V)$ are such that $D\widetilde{D}$ has retarded/advanced propagators $G^\pm_{{\scriptscriptstyle D} {\scriptscriptstyle\widetilde{D}}}$, then $D$ has retarded/advanced propagators
\[
G^{\pm}_\sDo = \widetilde D G^\pm_{{\scriptscriptstyle D} {\scriptscriptstyle\widetilde{D}}}.
\] 
\end{lemma}

\subsubsection{Phase space on Cauchy surface}\label{ssec:cauchy}

Let us fix a Cauchy surface $\Sigma$ of $(M,g)$. Consider a Green hyperbolic operator $D\in\Diff^m(M;V)$ (for the moment we do not assume it is formally self-adjoint). Let $V_\rho$ be a vector bundle over $\Sigma$ with a hermitian structure and let $\rho_\sDo :\Gamma_{\rm sc}(M;V)\to{\Gc}(\Sigma;V_\rho)$ be an operator which is the composition of a differential operator  with the pullback $\iota^\pback$ of the embedding $\iota:\Sigma\hookrightarrow M$.

We will say that $D$ is \emph{Cauchy hyperbolic} for the map $\rho_\sDo$ if the Cauchy problem
\beq\label{eq:cauchyD}
\begin{cases}Df=0, \quad f\in\Gamma_{\rm sc}(M;V)\\
\rho_\sDo f= \varphi,
\end{cases}
\eeq
has a unique solution for any initial datum $\varphi\in\Gc(\Sigma;V_\rho)$. 

In other words, the map $\rho_\sDo:\Ker D|_{\Gamma_{\rm sc}}\to{\Gc}(\Sigma;V_\rho)$ is a bijection. It can be proved that if $D$ is Green hyperbolic then there exists $\rho_\sD$ s.t. $D$ is Cauchy hyperbolic, see e.g. \cite[Sec. 4.4]{igorpreprint}.

By Cauchy hyperbolicity and Prop. \ref{prop:prebasic} there exists a unique operator $G_{\sDo\Sig}:{\Gc}(\Sigma;V_\rho)\to{\Gc}(\Sigma;V_\rho)$ s.t.
\[
G_\sDo= -G_\sDo \rho_\sDo^*  G_{\sDo\Sig} \rho_\sDo G_\sDo,
\]
where $\rho_\sDo^*$ is the formal adjoint of $\rho_\sDo$ w.r.t. the hermitian structures of $V$ and $V_\rho$. As a consequence of this definition,
\beq
\label{eq:idU}
\one=-G_\sDo\rho_\sDo^* G_{\sDo\Sig}\rho_\sDo \mbox{\ \ on \ } \Ker D|_{\Gsc}.
\eeq
This also implies $\rho_\sDo=-\rho_\sDo G_\sDo\rho_\sDo^* G_{\sDo\Sig}\rho_\sDo$ on $\Ker D|_{\Gsc}$, hence
\beq\label{eq:idU2}
\one=-\rho_\sDo G_{\sDo}\rho_\sDo^* G_{\sDo\Sig} \mbox{\ \ on \ } {\Gc}(\Sigma;V_\rho).
\eeq 
Note that (\ref{eq:idU2}) entails that $G_{\sDo\Sig}:{\Gc}(\Sigma;V_\rho)\to{\Gc}(\Sigma;V_\rho)$ is injective and by taking adjoints one can also show that it is surjective. 

It is useful to introduce the operator
\beq
U_{\sDo}\defeq -G_\sDo \rho_\sDo^* G_{\sDo\Sig}.
\eeq
By (\ref{eq:idU}) and (\ref{eq:idU2}), it satisfies $\rho_\sDo U_\sDo =\one$ and $U_\sDo \rho_\sDo = \one$ (on space-compact solutions of $D$). Moreover, $D U_\sDo=0$. Applying both sides of (\ref{eq:idU}) to $f$ we obtain that the solution of the Cauchy problem (\ref{eq:cauchyD}) is given by
\[
f=U_\sDo \varphi.
\]

If additionally $D=D^*$ then using (\ref{prop:basic2}) of Prop. \ref{prop:basic} we deduce that the phase space $(\cV_\sDo,q_\sDo)$ is isomorphic to $(\cV_{\sDo\Sig},q_{\sDo\Sig})$, which is defined in the following way:
\[
\cV_{\sDo\Sig}\defeq{\Gc}(\Sigma;V_\rho), \quad \bar u \,q_{\sDo\Sig} v\defeq \i (u| G_{\sDo\Sig} v)_{V_\rho}.
\]

\subsection{Gauge theory in subsidiary condition formalism}\label{ss:subsidiary}

In the setting proposed by Hack and Schenkel in \cite{HS}, the following data are used to define a classical linearized gauge field theory on a globally hyperbolic manifold $(M,g)$.

\begin{assumption}\label{as:subsidiary}Suppose that we are given:
\begin{enumerate}
\item bundles with hermitian structures $V_{0},V_{1}$ over $M$;
\item a formally self-adjoint operator $P\in\Diff(M;V_{1})$;
\item an operator $K\in\Diff(M;V_{0},V_{1})$, such that $K\neq0$ and 
\begin{enumerate}
\item $PK=0$,
\item $R\defeq K^* K\in\Diff(M;V_0)$ is Green hyperbolic;
\end{enumerate}
\item an operator $T\in\Diff(M;V_{0},V_{1})$, such that
\begin{enumerate}
\item $D\defeq P+TK^*\in\Diff(M;V_1)$ is Green hyperbolic;
\item $Q\defeq K^*T\in\Diff(M;V_0)$ is Green hyperbolic.
\end{enumerate}
\end{enumerate}
\end{assumption}

The operator $P$ accounts for the equations of motion, linearized around a background solution. The operator $K$ defines the linear gauge transformation $f\mapsto f+K g$, and the condition $PK=0$ states that $P$ is invariant under this transformation, which entails that $P$ is not hyperbolic. Making use of the assumption on $R$, the non-hyperbolic equation $Pf=0$ can always be reduced by gauge transformations to the subspace $K^* f=0$ of the hyperbolic problem $Df=0$. The equation $K^*f=0$ is traditionally called \emph{subsidiary condition} in the physics literature and can be thought as a covariant fixing of gauge. 

Let us first observe that the differential operators from Hypothesis \ref{as:subsidiary} satisfy the algebraic relations
\[
K^* D=QK^*, \quad DK=TR. 
\]
These have the following consequences on the level of propagators and spaces of solutions (statements (\ref{prop:notbasic1})--(\ref{longpropi4}) are proved in \cite{HS}).

\begin{proposition}\label{prop:notbasic}As a consequence of Hypothesis \ref{as:subsidiary},
\begin{enumerate}
\item\label{prop:notbasic1} $K^* G^{\pm}_{\sD}=G^{\pm}_\sQ K^*$ on ${\Gc}(M;V_1)$ and $K G^{\pm}_{\sR}=G^{\pm}_\sD T$ on ${\Gc}(M;V_0)$;
\item For all $\psi\in\Gsc(M;V_1)$ there exists $h\in\Gsc(M;V_0)$ s.t. $\psi-Kh\in\Ker K^*|_\Gsc$. If moreover $\psi\in\Ker P|_\Gsc$ then $\psi-K h\in \Ker P|_\Gsc \cap \Ker K^*|_\Gsc$;
\item\label{prop:notbasic3}We have 
\[
\Ker P|_\Gsc \cap \Ker K^*|_\Gsc\subset G_\sD \Ker K^*|_{\Gc} + G_\sD \Ran T|_{{\Gc}};
\]
\item\label{longpropi4} $\Ran P|_{{\Gc}}=\Ker K^*|_{{\Gc}}\cap G_\sD^{-1}\Ran K|_{\Gsc}$;
\item\label{longpropi5} $\Ran T|_{{\Gc}}\cap\Ker K^*|_{{\Gc}}=\{0\}$.
\end{enumerate}
\end{proposition}
\proof (\ref{longpropi5}): Suppose $u=Tf$ for $f\in\Gc$ and $K^*u=0$. Then $Qf=K^*T f=0$. But $Q$ is Green hyperbolic and hence has no compactly supported solutions.\qeds


In the subsidiary condition framework, the physical phase space associated to $P$, denoted $(\cV_\sP,q_\sP)$, is defined by
\[
\cV_\sP\defeq \frac{\Ker K^*|_{{\Gc}}}{\Ran P|_{{\Gc}}}, \quad \bar u \,q_\sP v\defeq \i (u| [G_\sD] v)_{V_1}.
\]

\begin{proposition}[\cite{HS}]The sesquilinear form $q_\sP$ is well defined on $\cV_\sP$.
\end{proposition}
\proof We need to show that $(u| G_\sD v)_{\sV_1}=0$ if $u\in\Ker K^*|_{\Gc}$ and $v=Pf$ for some $f\in{\Gc}(M;V_1)$. We have in such case
\[
G_\sD P f= - G_\sD T K^* f=-K G_\sR K^* f, 
\]
hence $(u|G_\sD P f)_{\sV_1}=-(K^* u | G_\sR K^* f)_{\sV_0}=0.$\qeds

It is possible to give different generalizations of Prop. \ref{prop:basic}, (\ref{prop:basic2}). Claim ${\rm a)}$ below is proved in \cite{HS}. {We prove that there is a different isomorphism (claim ${\rm b)}$) which is particularly useful as an intermediary step to find a `Cauchy surface version' of the phase space. It also formalizes the intuition that the phase space of space-compact solutions of $P$ should be equal $\Ker D|_{\Gsc}\cap\Ker K^*|_{\Gsc}$ modulo gauge transformations.}

\begin{proposition}\label{prop:Gpasses}The induced maps
\[
\begin{aligned} 
{\rm a)}\quad &[G_\sD]:\,\frac{\Ker K^*|_{{\Gc}}}{\Ran P|_{{\Gc}}}\longrightarrow \frac{\Ker P|_\Gsc}{\Ran K|_{\Gsc}},\\
{\rm b)}\quad &[G_\sD]:\,\frac{\Ker K^*|_{{\Gc}}}{\Ran P|_{{\Gc}}}\longrightarrow\frac{\Ker D|_{\Gsc}\cap\Ker K^*|_{\Gsc}}{\Ran G_\sD T |_{\Gc}},
\end{aligned}
\]
are both well defined and bijective. 
\end{proposition}
\proof b) \ For well-definedness we need to check that $f\in \Ker K^*|_{{\Gc}}$ implies $DG_\sD f=0$ (which is obvious) and $K^*G_\sD f=0$, which follows from $K^*G_\sD=G_\sQ K^*$.

For injectivity we need to show that if $u\in\Ker K^*|_{{\Gc}}$ and $G_\sD u\in\Ran G_\sD T|_{\Gc}$  then $u\in\Ran P|_{\Gc}$. By Prop. \ref{prop:notbasic}, (\ref{longpropi4}), it suffices to prove $G_\sD u \in \Ran K|_{\Gsc}$. Since $G_{\sD}T=K G_{\sR}$ we have $G_\sD u\in\Ran G_\sD T|_{\Gc} = \Ran K G_\sR |_{\Gc}$, which is contained in $\Ran K|_{\Gsc}$ as claimed.

Surjectivity amounts to showing
\[
\Ker D|_{\Gsc}\cap\Ker K^*|_{\Gsc}=G_\sD\Ker K^*|_{\Gc}+G_\sD\Ran T|_{\Gc}.
\]
The inclusion `$\supset$' is easy, the other one follows from Prop. \ref{prop:notbasic}, (\ref{prop:notbasic3}).\qed

\begin{remark}It is possible to construct directly a bijection\footnote{This remark is due to Christian G\'erard, private communication.}
\beq\label{eq:defI}
I : \, \frac{\Ker P|_\Gsc}{\Ran K|_{\Gsc}}\longrightarrow \frac{\Ker D|_{\Gsc}\cap\Ker K^*|_{\Gsc}}{\Ran G_\sD T |_{\Gc}}
\eeq
by setting for $\psi\in\Ker P|_\Gsc$
\[
I\psi\defeq \{ \psi-Kh : \ h \in \Gsc(M;V_0), \ Rh=K^*\psi \}.
\]
Using similar arguments as in \cite{HS}, one can show that $I\psi$ is not empty and 
\begin{itemize}
\item $I{\psi}+\Ran G_\sD T|_{\Gc}\subset I{\psi}$,
\item $\phi_1,\phi_2\in I\psi$ implies $\phi_1-\phi_2\in \Ran G_\sD T|_{\Gc}$, 
\item $I(\psi+Kf)=I\psi$ for all $\psi\in\Ker P|_\Gsc$, $f\in\Gsc(M;V_0)$.
\end{itemize}
These properties ensure that (\ref{eq:defI}) is well defined.
\end{remark}

\subsubsection{Phase spaces on a Cauchy surface}

To discuss the corresponding phase spaces on a fixed hypersurface $\Sigma\subset M$, we need to assume that the operators $D$ and $R$ are Cauchy-hyperbolic for some maps
\[
\rho_\sD: \ \Gamma(M;V_1)\to{\Gamma}(\Sigma;V_{\rho_\sD}), \quad \rho_\sR: \ \Gamma(M;V_0)\to{\Gamma}(\Sigma;V_{\rho_\sR}).
\]
We also need to have good analogues of the operators $K$ and $K^*$ on $\Sigma$.


Observe that $K$ maps solutions of $R$ to solutions of $D$, and $K^*$ maps solutions of $D$ to solutions of $Q$. Thus it makes sense to define 
\beq\label{eq:defKsig}
\begin{aligned}
K_\Sig&\defeq \rho_\sD K U_\sR: {\Gc}(\Sigma;V_{\rho_\sR})\to {\Gc}(\Sigma;V_{\rho_\sD}),\\
K^{\dagger}_{\Sig}&\defeq \rho_\sQ K^* U_\sD :  {\Gc}(\Sigma;V_{\rho_\sD})\to {\Gc}(\Sigma;V_{\rho_\sQ}).
\end{aligned}
\eeq
As in \ref{ssec:cauchy}, we associate to the Green hyperbolic operators $D,R,Q$ operators $G_{\sD\Sig}$, $U_{\sD}$, etc.

The notation $K^{\dagger}_{\Sig}$ is motivated by the fact that in the Maxwell and Yang-Mills case (where it is possible to choose $\rho_{\sR}=\rho_{\sQ}$), $K^{\dagger}_{\Sig}$ is the symplectic adjoint of $K_{\Sig}$, i.e. $K^{*}_{\Sig} G_{\sD\Sig} = G_{\sQ\Sig} K^{\dagger}_{\Sig}$, see \cite[Sec. 2.4]{GW2}. In general such relation does however not make sense because $K^{*}_{\Sig}$ and $K^{\dagger}_{\Sig}$ can have different target spaces. It also comes as a surprise that there is no need to consider a Cauchy version of the operator $T$ to get the phase space of Cauchy data.


\begin{lemma}\label{lem:cauchyrel}
\begin{enumerate}
\item\label{cauchyrelit1} $K U_\sR = U_\sD K_\Sig$ and $K^*U_\sD=U_\sQ K^{\dagger}_{\Sig}$;
\item\label{cauchyrelit0} $\rho_\sD K =K_\Sig \rho_\sR$ on $\Ker R|_\Gsc$ and $\rho_\sQ K^*=K^{\dagger}_{\Sig}\rho_\sD$ on $\Ker D|_\Gsc$;
\item\label{cauchyrelit2} $\Ker K^{\dagger}_{\Sig}|_{{\Gc}}=\rho_\sD G_\sD \Ker K^*|_{{\Gc}}$;
\item\label{cauchyrelit3} $\Ran K_{\Sig}|_{{\Gc}}=\rho_\sD G_\sD \Ran T|_{{\Gc}}$;
\item\label{cauchyrelit4} $K^{\dagger}_{\Sig} K_\Sig =0$.
\end{enumerate}
\end{lemma}
\proof (\ref{cauchyrelit1}), (\ref{cauchyrelit0}) and (\ref{cauchyrelit4}) follow easily from the definition of $K_\Sig$, $K^{\dagger}_{\Sig}$; (\ref{cauchyrelit2}) and (\ref{cauchyrelit3}) are proved as in \cite[Lem 2.9]{GW2}.\qed

\begin{proposition}\label{prop:rhopasses}The induced map
\[
[\rho_\sD]: \ \frac{\Ker D|_{\Gsc}\cap\Ker K^*|_{\Gsc}}{\Ran G_\sD T |_{\Gc}}\longrightarrow\frac{\Ker K_{\Sig}^{\dag}|_{\Gc}}{\Ran K_\Sig|_{\Gc}}
\]
is well defined and bijective.
\end{proposition}

\proof Recall that in the proof of Prop. \ref{prop:Gpasses}  we showed that $\Ker D|_{\Gsc}\cap\Ker K^*|_{\Gsc}=G_\sD\Ker K^*|_{\Gc}+G_\sD\Ran T|_{\Gc}$.

To show that $[\rho_\sD]$ is well defined and surjective it is thus sufficient to check that 
\[
\rho_\sD(G_\sD\Ker K^*|_{\Gc}+G_\sD\Ran T|_{\Gc})=\Ker K^{\dagger}_{\Sig}|_{\Gc},
\]
which  follows directly from (3), (4) and (5) of Lemma \ref{lem:cauchyrel}.

For injectivity we need to show that if $u\in G_\sD\Ker K^*|_{\Gc}+G_\sD\Ran T|_{\Gc}$ and $\rho_\sD u\in \Ran K_\Sig|_{{\Gc}}$ then $u\in\Ran G_\sD T|_{{\Gc}}$. This follows from (4) of Lemma \ref{lem:cauchyrel}.
\qeds

We deduce from Prop. \ref{prop:Gpasses} and Prop. \ref{prop:rhopasses} that  the map $\rho_\sD G_\sD$ induces an isomorphism between the phase space $(\cV_\sP,q_\sP)$ and the phase space $(\cV_{\Sig\sP},q_{\Sig\sP})$, defined in the following way:
\[
\cV_{\Sig\sP}\defeq\frac{\Ker K_{\Sig}^{\dag}|_{\Gc}}{\Ran K_\Sig|_{\Gc}}, \quad \bar u \,q_{\Sig\sP} v\defeq \i (u| [G_{\Sig\sD}] v)_{V_{{\rho}_\sR}}.
\]


\subsection{Gauge theory in abstract BRST formalism}\label{ss:BRST}

\subsubsection{The BRST framework at the linearized level}

\begin{definition}
A \emph{graded vector bundle} (indexed by a finite set $I\subset\zz$) is a direct sum of vector bundles $V=\bigoplus_{i\in I}V_{[i]}$, endowed with the corresponding grading.
\end{definition}

We identify sections of $V_{[i]}$ with corresponding sections of $V$ using the canonical embedding on fibers. By convention, if $i\notin I$ then $V_{[i]}$ is the zero bundle.



\begin{definition}
A graded vector bundle $V$ is \emph{hermitian}, if each $V_{[i]}$ is equipped with a hermitian structure. In such case we equip $V$ with the direct sum hermitian structure, denoted $(\cdot|\cdot)_V$.\end{definition}

Clearly, if $A$ decreases the grading by one, then its formal adjoint \wrt $(\cdot|\cdot)_V$, denoted $A^*$, increases it by one.

\begin{definition}\label{def:dif}
A \emph{differential} $A$ on a graded vector bundle $V$ is an operator $A \in \Diff(M; V)$ which fulfills
\begin{align*}
 A^2 & = 0, \\
 A \Gamma(M; V_{[i]}) & \subset \Gamma(M; V_{[i+1]}), \quad i\in I.
\end{align*}
Analogously, a \emph{codifferential} is nilpotent and decreases the grading.
\end{definition}

If $F\subset E\subset \Gamma(M;V)$ we write
\[
E/F {\big|_{[i]} }\defeq \frac{E \cap\Gamma(M;V_{[i]})}{F \cap \Gamma(M;V_{[i]})}.
\]

The outcome of the BRST method can be put in an abstract framework as follows. 

\begin{assumption}\label{asBRST}Suppose that we are given:
\begin{enumerate}
\item a hermitian graded vector bundle $V$ over $M$ (we denote the grading by $\gh$);
\item a codifferential $\gamma\in \Diff(M;V)$ s.t.
\[
\begin{aligned}
H_{-1,\c}(\gamma)\defeq\frac{\Ker \gamma|_{\Gc}}{\Ran \gamma|_{\Gc}} \,{\bigg|_{[-1]}}=\{0\};
\end{aligned}
\]
\item a Green hyperbolic operator $L\in\Diff(M;V)$, s.t. $L=L^*$ and
\[
\begin{aligned}
{\rm a)}& \quad \gamma^*L=L\gamma \\
{\rm b)}&\quad  L  \Gamma(M; V_{[i]}) \subset \Gamma(M; V_{[-i]}), \quad i\in I = \{ -1,0,1 \}.
\end{aligned}
\]
\end{enumerate}
\end{assumption}

The operator $\gamma$ is the formal adjoint of the BRST differential, which generates the BRST symmetry. 

We postulate that the classical phase space in the BRST framework associated to the data in Hypothesis \ref{asBRST} is $(\cV,q)$, where
\beq\label{eq:defphsp}
\cV\defeq\frac{\Ker \gamma^*|_{{\Gc}}}{\left(\Ran \gamma^*|_{{\Gc}}+ (\Ran L|_{{\Gc}}\cap \Ker \gamma^*|_{{\Gc}} ) \right)}\Gho\,, \quad {\overline{u}} \,q  v\defeq \i ( u| [G_\sL] v)_V
\eeq

\begin{lemma}\label{lem:gamgl}If  Hypothesis \ref{asBRST} holds then $ G^\pm_\sL \gamma^*= \gamma G^\pm_\sL $.
\end{lemma}

\begin{proposition}\label{eq:checkq}The sesquilinear form $q$ is well defined on $\cV$.\end{proposition}
\proof It suffices to check that $(u|G_{\sL} v)_{V}=0$ if $u\in\Ker \gamma^*|_{{\Gc}}$ and $v=\gamma^*f+Lh$. We have indeed in such case 
\[
(u|G_{\sL} v)_{V}=(u|G_{\sL} (\gamma^*f+Lh))_{V}=(\gamma^*u|G_{\sL} f)_{V}=0.\qed
\]

Our definition (\ref{eq:defphsp}) of the phase space is justified by the next proposition, which relates it to a subspace of space-compact solutions of $L$ and to $\gamma$-cohomology at the same time. The proof relies in an essential way on all parts of Assumption \ref{asBRST}.

\begin{proposition}\label{prop:induced1}The induced map
\[
[G_{\sL}] : \ \frac{\Ker \gamma^*|_{{\Gc}}}{\left(\Ran \gamma^*|_{{\Gc}}+\Ran L|_{{\Gc}}\cap \Ker \gamma^*|_{{\Gc}}\right)} \Gho\,\longrightarrow \frac{\Ker L|_{\Gsc}\cap \Ker \gamma|_{\Gsc}}{\Ran G_\sL\gamma^*|_{\Gc}}\Gho
\]
is well defined and bijective.
\end{proposition}
\proof Observe that by Hypothesis~\ref{asBRST} (3b), $G_\sL$ preserves the subspace of $\gh = 0$. Thus for well-definedness it suffices to show the inclusions 
\[
\begin{aligned}
&G_\sL\Ker \gamma^*|_{{\Gc}}\subset \Ker L|_{\Gsc}\cap \Ker \gamma|_{\Gsc}, \\ &G_{\sL}(\Ran \gamma^*|_{{\Gc}}+\Ran L|_{{\Gc}})\subset\Ran G_\sL\gamma^*|_\Gsc,
\end{aligned}
\]
which are both straightforward to check.

For injectivity it suffices to show that if $u\in\Ker \gamma^*|_{{\Gc}}$ and $G_\sL u = G_\sL \gamma^* f$ for some $f\in{\Gc}$ then $u\in\Ran \gamma^*|_{{\Gc}}+(\Ran L|_{{\Gc}}\cap \Ker \gamma^*|_{{\Gc}})$. By $G_\sL (u-\gamma^*f)=0$ there exists $k$ s.t. $u-\gamma^* f=Lk$, hence $u=\gamma^*f + Lk$ as requested. 

Surjectivity amounts to 
\beq\label{eq:prsurj}
\big(\Ker L|_{\Gsc}\cap \Ker \gamma|_{\Gsc}\big){\big|_{[0]}}\,\subset \big( G_{\sL}\Ker \gamma^*|_{{\Gc}}\big){\big|_{[0]}}.
\eeq
To prove this, observe that if $\psi\in(\Ker L|_{\Gsc})\gho$ then there exists $h\in\Gc(M;V_{[0]})$ s.t. $\psi=G_\sL h$ and if additionally $\psi\in \Ker \gamma|_{\Gsc}$ then $G_\sL \gamma^* h=\gamma G_\sL h=0$. This implies 
\beq\label{eq:prsurj2}
\gamma^* h=L k
\eeq
for some $k\in\Gc(M;V)$. By (2) and (3) b) of Hypothesis \ref{asBRST}, $k$ belongs to $\Gc(M;V_{[-1]})$. Moreover, (\ref{eq:prsurj2}) implies
\[
\gamma k = \gamma G^+_\sL L k =   G^+_\sL \gamma^* L k = G^+_\sL (\gamma^*)^2  h = 0.
\]
By (2) of Hypothesis \ref{asBRST} this implies $k=\gamma \tilde k$ for some $\tilde k\in\Gc(M;V)$. It follows that $\psi=G_\sL(h-L\tilde{k})$ with 
\[
\gamma^*(h-L\tilde{k})=\gamma^* h-L \gamma \tilde{k}=\gamma^* h - L k=0.
\]
This proves (\ref{eq:prsurj}).\qed

\subsubsection{Phase spaces on Cauchy surface}\label{sss:phcb}

We now discuss the phase spaces on a Cauchy surface $\Sigma\subset M$.


Since $L$ is Green hyperbolic, there exists a hermitian vector bundle $V_{\rho}$ over $\Sigma$ and a map
\[
\rho: \Gsc(M;V)\to{\Gc}(\Sigma;V_{\rho})
\]
such that $L$ is Cauchy hyperbolic for $\rho$.
We equip $V_\rho$ with the grading inherited from $V$, also denoted $\gh$.


We define an analogue of the operator $\gamma$, acting on Cauchy data:
\[
\gamma_\Sig\defeq \rho \gamma U_\sL: {\Gc}(\Sigma;V_{\rho})\to{\Gc}(\Sigma;V_{\rho}).
\]
This operator decreases the grading, by the compatibility of $\gamma$ with $L$.

\begin{lemma}\label{gotocbrst}Let $\gamma_\Sig$ be defined above. Then:
\begin{enumerate}
\item $\gamma U_\sL = U_\sL \gamma_\Sig$ on ${\Gc}(\Sigma;V_{\rho})$ and $\gamma_{\Sig}\rho=\rho \gamma$ on $\Ker L|_\Gsc$;
\item $\Ker \gamma_{\Sig}|_{{\Gc}}=\rho (\Ker L|_\Gsc\cap\Ker \gamma|_\Gsc)$;
\item $\Ran \gamma_{\Sig}|_{{\Gc}}=\rho \Ran G_\sL \gamma^*|_{{\Gc}}$;
\item $\gamma_\Sig^2=0$;
\item $\gamma_\Sig^\dag=\gamma_\Sig$, i.e. $\gamma_\Sig^* G_{\Sig\sL} = G_{\Sig\sL} \gamma_\Sig$.
\end{enumerate}
\end{lemma}
\proof (1) \& (4): These follow easily from the definition of $\gamma_\Sig$ and the identities $U_\sL \rho=\one$ on $\Ker L|_\Gsc$ and $\rho U_\sL =\one$.


(2): If $u=\rho f$ with $f\in\Ker L|_\Gsc\cap\Ker \gamma|_\Gsc$ then $\gamma_{\Sig}\rho f=\rho\gamma f=0$. Conversely, if $u\in\Ker \gamma_{\Sig}|_{{\Gc}}$ then using that $\one=\rho U_{\sL}$ we get $u=\rho f$ with $f=U_\sL u$ and by (1)
\[
\gamma f =\gamma U_\sL u = U_\sL \gamma_{\Sig} u=0, \quad L f = L U_\sL u = 0.
\]

(3): If $u=\rho G_\sL \gamma^* f$ then $u=\rho \gamma G_\sL f=\gamma_{\Sig}\rho G_\sL f$. Conversely, if $u=\gamma_{\Sig} h$ then using that $\one=-\rho G_\sL \rho^* G_{\sL\Sig}$ we get 
\[
u=-\rho G_\sL \rho^* G_{\sL\Sig} \gamma_{\Sig} h =-\rho G_\sL \gamma^* \rho^* G_{\sL\Sig} h.
\]

(5): Using (1) and $G_{\sL\Sig}^*=-G_{\sL\Sig}$  we compute
\[
\begin{aligned}
\gamma_\Sig^* G_{\sL\Sig} & = U_\sL^*  \gamma^* \rho^* G_{\sL\Sig}=G_{\sL\Sig}\rho G_\sL^* \gamma^* \rho^* G_{\sL\Sig} \\ & =G_{\sL\Sig} \gamma_\Sig \rho G_\sL^* \rho^* G_{\sL\Sig}= G_{\sL\Sig} \gamma_\Sig \rho U_\sL  = G_{\sL\Sig} \gamma_\Sig.\qeds
\end{aligned}
\]

Since $\rho$ preserves $\gh$, as a corollary of Lemma \ref{gotocbrst} we obtain the following result. 

\begin{proposition}\label{prop:rhopassesbrs}The induced map
\[
[\rho]: \ \frac{\Ker L|_{\Gsc}\cap \Ker \gamma|_{\Gsc}}{\Ran G_\sL\gamma^*|_{\Gc}}\,\Gho\,\longrightarrow\frac{\Ker \gamma_{\Sig}|_{\Gc}}{\Ran \gamma_{\Sig}|_{\Gc}}\,\Gho
\]
is well defined and bijective.
\end{proposition}

We deduce from Prop. \ref{prop:induced1} and Prop. \ref{prop:rhopassesbrs} that the map $\rho G_\sL$ induces an isomorphism between the phase space $(\cV,q)$ and the phase space $(\cV_\Sig,q_\Sig)$, defined in the following way:
\beq\label{eq:defphys}
\cV_\Sig\defeq\frac{\Ker \gamma_{\Sig}|_{\Gc}}{\Ran \gamma_{\Sig}|_{\Gc}}\,\Gho\,, \quad \bar u \,q_\Sig v\defeq \i (u| [G_{\sL\Sig}] v)_{V_{\rho}}.
\eeq

\subsection{Non-degeneracy of the phase space}\label{ss:nond}

In what follows we will formulate a criterion for non-degeneracy of the phase space $\cV_\Sig$ in terms of BRST cohomology. The obvious advantage of working with Cauchy data is that $\cV_\Sig$ is given by a much simpler formula than $\cV$, namely it involves only one operator $\gamma_\Sig$ (which is even a differential operator).

\subsubsection{Notations --- cohomology}\label{sss:coh}

Let $A$ be a differential on a graded vector bundle $V$ over $\Sigma$.  We introduce the \emph{smooth}, resp.\ \emph{compactly supported $A$-cohomology} of $\Sigma$:
\[
H^i(A)\defeq \frac{\Ker A|_{\Gamma}}{\Ran A|_{\Gamma}}\,\Ghi\,, \quad H^i_{\rm c}(A)\defeq \frac{\Ker A|_{\Gc}}{\Ran A|_{\Gc}}\,\Ghi\,.
\]
For a codifferential $B$, one analogously defines the \emph{homologies} $H_i(B)$ and $H_{i, \c}(B)$.
\subsubsection{Non-degeneracy criteria}

The embedding of $\Ker \gamma^*_{\Sig}|_{\Gc}$ into $\Ker \gamma^*_{\Sig}|_{\Gamma}$ and the embedding of $\Ker \gamma^*_{\Sig}|_{\Gamma}$ into the space 
\[
\{ u\in \Gc' :  (u|v )_{V_\rho}=0 \ \forall\,v\in\Ran\gamma_{\Sig}|_{\Gamma_{\rm c}}\}
\]
 induce maps on the respective cohomologies, denoted
\beq\label{eq:coh1}
H^0_{\rm c}(\gamma^*_\Sig) \stackrel{\imath}{\longrightarrow} H^0(\gamma^*_\Sig) \stackrel{\jmath}{\longrightarrow} \big(H_{0, \rm c}(\gamma_\Sig) \big)^*. 
\eeq
It turns out that the issue of (non)-degeneracy of $q_\Sig$ is directly related to injectivity of the maps in (\ref{eq:coh1}). For instance, injectivity of $\imath$ reads
\beq\label{eq:jinj}
\Ran\gamma_\Sig^*|_{\Gc} = \big(\Ran\gamma_\Sig^*|_{\Gamma}\big) \cap \Gamma_{\rm c}(\Sigma;V_\rho) 
\eeq
on ghost number zero sections. This condition can be thought as the BRST analogue of the criterion stated in \cite[Prop.~3.5]{SDH} for the Maxwell field in the subsidiary condition framework. 

On the other hand, injectivity of $\jmath$ amounts to
\beq\label{eq:jjjjjinj}
\Ker\gamma_\Sig^*|_{\Gamma}\cap(\Ker\gamma_{\Sig}|_{\Gc})^*=\Ran\gamma_\Sig^*|_{\Gamma}.
\eeq
We will see later on that in the case of Maxwell fields, this is a trivial consequence of Poincar\'e duality, we will thus term property (\ref{eq:jjjjjinj}) \emph{generalized Poincar\'e duality} in the generic case. Assuming that the generalized Poincar\'e duality holds true, non-degeneracy of $q_\Sig$ can be conveniently studied in terms of injectivity of $\imath$.


\begin{theoreme}\label{thm:criterion}
Let $q_\Sig$ be defined in (\ref{eq:defphys}). In terms of the maps defined in (\ref{eq:coh1}):
\begin{enumerate}
\item $q_\Sig$ is non-degenerate on $\cV_\Sig= H_{0, \rm c}(\gamma_\Sig)$ iff $\jmath\circ\imath$ is injective.
\item If $\imath$ is not injective then $q_\Sig$ is degenerate.
\item Suppose $\jmath$ is injective. Then $q_\Sig$ is non-degenerate iff $\imath$ is injective. 
\end{enumerate}
\end{theoreme}
\proof (1): \ For simplicity of notation we drop the $\gho$ subscripts. Non-degeneracy of $q_\Sig$ on $\cV_\Sig$ is equivalent to the property that for any $u\in\Ker \gamma_\Sig|_{\Gamma_\c}$:
\beq\label{eq:tscpt}
\big(\,(f|G_{\sL \Sig} u)_{V_{\rho}}=0 \quad \forall\,f\in\Ker \gamma_\Sig|_{\Gamma_\c}\,\big)\,\Longleftrightarrow\,\big(u\in\Ran \gamma_\Sig|_{\Gamma_\c}\big).
\eeq
But since $G_{\sL\Sig}$ is bijective on $\Gamma_{\rm c}$ and $G_{\sL\Sig}\gamma_\Sig=\gamma_{\Sig}^* G_{\sL\Sig}$ (see Lemma \ref{gotocbrst}), the \rhs of (\ref{eq:tscpt})  is equivalent to $g\defeq G_{\sL\Sig}u\in\Ran \gamma^*_\Sig|_{\Gamma_{\rm c}}$. Hence, (\ref{eq:tscpt}) holds true iff $\jmath\circ\imath$ is injective.

(2) \& (3): \ This follows from (1).
\qeds

As a straightforward corollary we obtain that if $\jmath$ is injective and the Cauchy surface $\Sigma$ is compact then $q_\Sig$ is non-degenerate. Indeed, injectivity of $\imath$ (i.e. (\ref{eq:jinj})) is in such case automatically satisfied.  

\begin{remark}\label{rem:rem}From the proof of Thm. \ref{thm:criterion} one sees that if $\imath$ is not injective then actually \emph{any} hermitian form of the form $(\cdot|\lambda_\Sig \cdot)_{V_\rho}$ is degenerate, supposing $\lambda_\Sig:\Gamma_\c\to\Gamma$ satisfies $\lambda_\Sig^*=\lambda_\Sig$, $\lambda_\Sig \gamma_\Sig = \gamma^*_\Sig \lambda_\Sig$.
\end{remark}


\subsection{Relation between the two frameworks}\label{ss:relation}

In this section we discuss the relation between the BRST formalism in our setup and the subsidiary condition framework. 

We first introduce a modified set of assumptions that describes more accurately some of the examples met in the literature.

\begin{assumption}\label{as:subsidiary2}Suppose that we are given:
\begin{enumerate}
\item bundles with hermitian structures $V_{0},V_{1}$ over $M$;
\item $P\in\Diff(M;V_{1})$ s.t. $P^*=P$;
\item an operator $K\in\Diff(M;V_{0},V_{1})$, such that $K\neq0$ and 
\begin{enumerate}
\item $PK=0$,
\item the operator
\[
L\defeq
 \begin{pmatrix}
   P &  K & 0 & 0 \\
   K^* & -\alpha\one & 0 & 0 \\
   0 & 0 & 0 & K^*K \\
   0 & 0 & K^*K & 0  \end{pmatrix} \in \Diff(M;V_1\oplus V_0^{\oplus 3})
\]  
\end{enumerate}
is Green hyperbolic for some $\alpha\in\rr$.
\end{enumerate}
\end{assumption}

We show that the subsidiary condition framework of \cite{HS} with $K=T$ is a special case of the above assumptions.

\begin{lemma}\label{lem:rrel2}Suppose $P$ and $K$ satisfy Hypothesis \ref{as:subsidiary} with $K=T$, in particalar
$D\defeq P+KK^*$ and $Q\defeq K^*K$ are Green hyperbolic. Then Hypothesis \ref{as:subsidiary2} is satisfied for arbitrary $\alpha$.
\end{lemma}
\proof To prove that $L$ is Green-hyperbolic, observe that the operators
\beq\label{eq:GL}
G_\sL^\pm \defeq \begin{pmatrix}

 G_\sD^\pm (\one + (\alpha-1) K K^* G_\sD^\pm) & K G_\sQ^\pm &  0 & 0 \\
 K^* G_\sD^\pm & 0 & 0 & 0 \\
 0 & 0 & 0 & G_\sQ^\pm \\
 0 & 0 & G_\sQ^\pm & 0
\end{pmatrix}
\eeq
satisfy $L G_\sL^\pm = G_\sL^\pm L=\one$ (here one uses $K^* G^\pm_\sD K = \one$, as a consequence of Proposition~\ref{prop:notbasic} (1)) and have the support properties required for advanced, resp. retarded propagators.\qeds


Let us set $V\defeq V_1\oplus (V_0)^{\oplus 3}$ and
\beq\label{eq:defgam}
   \gamma \defeq \begin{pmatrix}
 0 & 0 &  K & 0 \\
 0 & 0 & 0 & 0 \\
 0 & 0 & 0 & 0 \\
 0 & \one & 0 & 0
\end{pmatrix}\in\Diff(M;V_1\oplus (V_0)^{\oplus 3}).
\eeq
We have obviously $\gamma^2=0$. We equip the bundle $V$ with the obvious hermitian structure
\[
( f | g)_{V}\defeq (f_\sA|g_\sA)_{V_1} + (f_\sB|g_\sB)_{V_0} + (f_c|g_{c})_{V_0} + (f_{\bar{c}}|g_{\bar c})_{V_0} 
\]
for $f=(f_\sA,f_\sB,f_c,f_{\bar c})$, $g=(g_\sA,g_\sB,g_c,g_{\bar c})\in{\Gc}(M;V_1\oplus (V_0)^{\oplus 3})$.

We also equip $V$ with a grading $\gh$, which can be written symbolically as
\[
\gh\defeq \begin{pmatrix}
 0 & 0 &  0 & 0 \\
 0 & 0 & 0 & 0 \\
 0 & 0 & 1 & 0 \\
 0 & 0 & 0 & -1
\end{pmatrix}
\]
or in other terms $V=V_{[0]}\oplus V_{[1]}\oplus V_{[-1]}$, where $V_{[0]}=V_1\oplus V_0$, $V_{[1]}=V_0$ and $V_{[-1]}=V_0$. This way, $\gamma$ is a codifferential in the sense of Def. \ref{def:dif}.


\begin{proposition}The operators $L$, $\gamma$ satisfy the assumptions of the BRST framework (Hypothesis \ref{asBRST}).
\end{proposition}
\proof The identity $\gamma^* L = L \gamma$ and the property of preserving/decreasing $\gh$ are straightforward to check. Furthermore, we compute (skipping `$|_{{\Gc}}$' in the notation):
\[
\begin{aligned}
\Ker \gamma&=\Gc(M;V_1)\oplus  \{0\} \oplus \Ker K \oplus  {\Gc}(M;V_{[-1]}),\\
\Ran \gamma&=\Ran K \oplus \{0\} \oplus   \{0\} \oplus {\Gc}(M;V_{[-1]}).
\end{aligned}
\]
We thus see that the homology of $\gamma$ at ghost number $-1$ is trivial.\qeds

The formal adjoint of $\gamma$ wrt. $(\cdot|\cdot)_{V}$ is
\[
\gamma^* = \begin{pmatrix}
 0 & 0 &  0 & 0 \\
 0 & 0 & 0 & \one \\
 K^* & 0 & 0 & 0 \\
 0 & 0 & 0 & 0
\end{pmatrix}
\in\Diff(M;V_1\oplus (V_0)^{\oplus 3}).
\]

We compute (skipping `$|_{{\Gc}}$' in the notation):
\[
\begin{aligned}
&\Ker \gamma^*= \Ker K^* \oplus {\Gc}(M;V_0) \oplus {\Gc}(M;V_0) \oplus \{0\},\\
&\Ran \gamma^*=\{0\}\oplus {\Gc}(M;V_0) \oplus \Ran K^* \oplus \{0\},\\
&\Ran L=(\Ran P+\Ran K)\oplus (\Ran K^* -\alpha{\Gc}(M;V_0)) \oplus \Ran Q\oplus \Ran Q,
\end{aligned}
\]
where $Q=K^*K$, therefore 
\[
\begin{aligned}
\Ran \gamma^*+(\Ran L\cap\Ker \gamma^*)=&\,(\Ran P+\Ran K)\cap\Ker K^*\oplus {\Gc}(M;V_0) \\
  & \oplus(\Ran K^*+\Ran Q) \oplus\{0\}\\
=&\, \Ran P\oplus {\Gc}(M;V_0) \oplus \Ran K^* \oplus \{0\},
\end{aligned}
\]
where we used that $\Ran Q\subset\Ran K^*$, $\Ran P\subset\Ker K^*$ and $\Ran K\cap\Ker K^*=\{0\}$ for compactly supported sections (the last fact is proved as (\ref{longpropi5}) of Prop. \ref{prop:notbasic}). It follows that
\beq\label{eq:expressforalsp}
\frac{\Ker \gamma^*}{\Ran \gamma^*+(\Ran L\cap \Ker \gamma^*)}=\underbrace{\frac{\Ker K^*}{\Ran P}\oplus\{0\}}_{[0]}\oplus\underbrace{\frac{{\Gc}(M;V_0)}{\Ran K^*}}_{[1]}\oplus\underbrace{\{0\}}_{[-1]}.
\eeq
To relate the symplectic forms, we compute, using (\ref{eq:GL}), and for $f=(f_\sA,f_\sB,f_c,f_{\bar c})$ and $g=(g_\sA,g_\sB,g_c,g_{\bar c})\in{\Gc}(M;V_1\oplus (V_0)^{\oplus 3})$,
\[
\begin{aligned}
(f|G_{\sL} g)_{\sVA}=&\,(f_{\sA}| G_\sD g_\sA)_{\sV_1} + (\alpha-1) (f_\sA| G' g_\sA)_{\sV_1} +(f_{\sA}|K G_\sQ g_\sB)_{\sV_1} \\ &+ (f_\sB|K^* G_\sD g_\sA)_{\sV_0} + (f_{c}| G_\sQ g_{\bar c})_{\sV_0} + (f_{\bar c}| G_\sQ g_c)_{\sV_0},
\end{aligned}
\]
where
\[
 G' = G_\sD^+ K G_\sQ^+ K^* - G_\sD^- K G_\sQ^- K^*.
\]
Hence, for $f,g\in\Ker \gamma^*|_{{\Gc}(M; V_{[0]})}$ we simply have $\overline{f} q g = \overline{f_\sA} q_{\sP} g_\sA$.

We conclude that the phase space $(\cV,q)$ in the BRST framework and $(\cV_\sP,q_\sP)$ in the subsidiary condition framework are in this case isomorphic (i.e. when $T=K$). In the case $T\neq K$ it is in general not clear how to construct the operator $L$, we will see several possible choices for the Rarita-Schwinger equation in Subsect. \ref{ss:RS}.  

\subsubsection{Phases spaces on a hypersurface}

We can also directly compare the Cauchy surface phase spaces $(\cV,q)$, $(\cV_\sP,q_\sP)$.

Let us assume Hypothesis \ref{as:subsidiary} with $K=T$, so that by Lemma \ref{lem:rrel2}, Hypothesis \ref{as:subsidiary2} is satisfied with $\alpha=1$, to which we restrict in the following. This also entails that $D$ and $Q$ are Cauchy-hyperbolic for some $\rho_\sD$, $\rho_\sQ$.  

Observe that the equation $Lf=0$ for $f=(f_\sA,f_\sB,f_c,f_{\bar c})\in{\Gc}(M;V_1\oplus (V_0)^{\oplus 3})$ is equivalent to
\[
\begin{cases} D f_\sA=0, \\
K^* f_\sA = f_\sB, \\
Q f_c = Q f_{\bar c}=0.
\end{cases}
\]
It follows that $L$ is Cauchy hyperbolic for the map
\[
\rho f \defeq (\rho_\sD f_\sA, \rho_\sQ f_c, \rho_\sQ f_{\bar c}).
\] 
Moreover, $\gamma_{\Sig} \rho = \rho \gamma$ on $\Ker L|_\Gsc$ for
\[
\gamma_{\Sig} = \begin{pmatrix}
 0 & K_\Sig & 0 \\
 0 & 0 & 0 \\
 K_{\Sig}^{\dag} & 0 & 0 
\end{pmatrix}\in\Diff(\Sigma; V_{\rho_\sD}\oplus (V_{\rho_\sQ})^{\oplus2}).
\]
We compute
\[
\begin{aligned}
\Ker \gamma_{\Sig}|_{{\Gc}}&=\Ker K_{\Sig}^{\dag}|_{{\Gc}}\oplus \Ker K_{\Sig}|_{{\Gc}} \oplus \Gc(\Sigma;V_{\rho_\sQ})\\
\Ran \gamma_{\Sig}|_{{\Gc}}&=\Ran K_{\Sig}|_{{\Gc}}\oplus \{0\} \oplus \Ran K_{\Sig}^{\dag}|_{{\Gc}},
\end{aligned}
\]
hence
\[
\frac{\Ker \gamma_{\Sig}|_{{\Gc}}}{\Ran \gamma_{\Sig}|_{{\Gc}}}=\underbrace{\frac{\Ker K_{\Sig}^{\dag}|_{{\Gc}}}{\Ran K_{\Sig}|_{{\Gc}}}}_{[0]} \oplus\underbrace{ \Ker K_{\Sig}|_{{\Gc}} }_{[1]}\oplus\underbrace{\frac{\Gc(\Sigma;V_{\rho_\sR})}{\Ran K_\Sig^\dag|_{\Gc}}}_{[-1]}.
\]

\section{Hadamard states}\label{sec:states}

\subsection{Quasi-free states}

Let $(\cV,q)$ be a phase space (i.e. $\cV$ is a complex vector space and $q$ a hermitian form on $\cV$). 

We denote $\aCCR(\cV,q)$ the associated {\em polynomial  CCR $*$-algebra} (see eg. \cite[Sect. 8.3.1]{derger}), and (if $q\geq 0$) $\aCAR(\cV,q)$ the {\em polynomial  CAR $*$-algebra}. Recall that $\aCCR(\cV,q)$ is generated by elements $\psi(v)$, $\psi^*(w)$ (the \emph{abstract complex field operators}) subject to commutation relations
\beq\label{eq:CCR}
[\psi(v), \psi(w)]= [\psi^{*}(v), \psi^{*}(w)]=0,  \ \ [\psi(v), \psi^{*}(w)]=  \bar{v} q w \one, \ \ v, w\in \cV,
\eeq
whereas $\aCAR(\cV,q)$ is generated by elements satisfying analogous anti-com\-mu\-tation relations. More precisely, the assignment $v\mapsto\psi(v)$ is anti-$\cc$-linear, whereas $v\mapsto\psi^*(v)$ is $\cc$-linear, see e.g. \cite{wrothesis,GW} for the transition to the more commonly used real vector space terminology.

The \emph{complex covariances} of a state $\omega$ on $\aCCR(\cV,q)$ or $\aCAR(\cV,q)$ are defined by
\[
\bar{v}\Lambda^+ w \defeq \omega\big(\psi(v)\psi^*(w)\big), \quad \bar{v}\Lambda^- w \defeq \omega\big(\psi^*(w)\psi(v)\big), \quad v,w\in \cV.
\]
It is well known that two hermitian forms $\Lambda^\pm$ on $\cV$ are the complex covariances of a \emph{quasi-free, gauge-invariant} state on $\aCCR(\cV,q)$, resp. $\aCAR(\cV,q)$ iff
\[
\Lambda^\pm \geq 0, \quad \Lambda^+-\Lambda^-=q,
\]
respectively
\[
\Lambda^\pm \geq 0, \quad \Lambda^++\Lambda^-=q,
\]
see for instance \cite{wrothesis} and references therein. 

\subsection{Hadamard two-point functions}





Let $V$ be a graded vector bundle (the grading is denoted $\gh$) and let $L\in\Diff(M;V)$ be Green hyperbolic.

Let us denote symbolically $(-1)^{\rm gh}$ the matrix with entries $(-1)^{ij}$, where the indices $i,j$ refer to the grading of $V=\bigoplus_{i\in I}V_{[i]}$. 

We say that a pair of operators $\lambda^\pm_\sL : \Gamma_{\rm c}(M; V)\to \Gamma_{\rm c}'(M; V)$ are \emph{bosonic}, resp. \emph{fermionic} \emph{two-point functions for $L$} if
\beq \label{eq:def2p}
\begin{aligned}
i)&\quad \lambda^{\pm}_\sL : \Gamma_{\rm c}(M; V)\to \Gamma(M; V)\\[1mm]
ii)&\quad \lambda^{\pm}_\sL = \lambda^{\pm*}_\sL  \hbox{ for }(\cdot| \cdot)_{V} \hbox{ on }\Gamma_{\rm c}(M; V),\\[1mm]
iii)&\quad \lambda^{\pm}_\sL L=0,\\[1mm]
iv)&\quad \lambda^{\pm}_\sL \Gc(M ; V_{[i]}) \subset \Gc(M ; V_{[-i]}), \quad i\in I, \\[1mm]
v)&\quad \lambda^{+}_\sL \mp(-1)^{\rm gh}\lambda^{-}_\sL = \i G_\sL, \\[1mm] 
\end{aligned}
\eeq
where in the last equation the sign `$-$' corresponds to the bosonic case, and the `$+$' sign to the fermionic case. 

At this stage we have not imposed any positivity condition on $\lambda^\pm_\sL$, so in our terminology a pair of two-point function does not have to correspond to complex covariances of a state. 


We say that a pair of (bosonic, fermionic) two-point functions $\lambda^\pm_\sL$ is \emph{Hadamard} if
\[
\begin{aligned}
\musc\quad&\wf'(\lambda^\pm_\sL)= (\cN^\pm\times\cN^\pm)\cap \wf'(G_\sL),
\end{aligned}
\]
where 
\[
\cN^\pm\defeq \{(x,\xi)\in T_x^* M\setminus\{0\} : \ g^{\mu\nu}(x)\xi_\mu \xi_\nu =0, \ \xi\in V_x^{\pm*}\},
\]  
and $V_x^{\pm*}$ are the positive/negative energy cones above $x\in M$ (see  \cite{H1} for the definition of the primed wave front set $\wf'$). This form of the Hadamard condition is equivalent to the one used originally by Radzikowski \cite{radzikowski}, this is also equivalent to the condition
\beq\label{eq:hadasv}
\wf'(\lambda^\pm_\sL)\subset \cN^\pm\times\cN^\pm,
\eeq
which appeared first in \cite{SV,morehollands}, see \cite{wrothesis} for a review on this topic.
\medskip

Let us now assume $L,\gamma\in\Diff(M;V)$ satisfy the assumptions of the BRST formalism (Hypothesis \ref{asBRST}), and let $(\cV,q)$ be the associated phase space (defined in (\ref{eq:defphsp})). 

\begin{definition}\label{def:hadaa}We say that a bosonic (fermionic) quasi-free state $\omega$ on $\aCCR(\cV,q)$ ($\aCAR(\cV,q)$) is \emph{Hadamard} if  there exists Hadamard bosonic (fermionic) two-point functions $\lambda_\sL^{\pm}$ for $L$, s.t. the complex covariances $\Lambda^\pm$ of $\omega$ are given by:
\beq\label{eq:fromtpftoc}
\overline{[u]}\Lambda^{\pm}[v]=  (u| \lambda_\sL^{\pm}v)_{V}, \ \ u, v\in \Ker \gamma^{*}|_{\Gamma_{\rm c}(M;V_{[0]})},
\eeq
where 
\[
\Ker \gamma^{*}|_{\Gamma_{\rm c}(M;V_{[0]})}\ni u\mapsto [u]\in  \frac{\Ker \gamma^*|_{{\Gc}}}{\left(\Ran \gamma^*|_{{\Gc}}+\Ran L|_{{\Gc}}\cap \Ker \gamma^*|_{{\Gc}}\right)}\Gho\,=\cV
\]
is the canonical map.
\end{definition}
We say that $\lambda_{\sL}^{\pm}$ are the two-point functions of the Hadamard state $\omega$.

The next lemma is an analogue of \cite[Lemma 3.16]{GW2} in the BRST formalism and gives a more practical characterization of two-point functions.

\begin{lemma}\label{lem:obvious}
 $\lambda_\sL^{\pm}: \Gamma_{\rm c}(M; V)\to \Gamma(M; V)$ are the two-point functions of a Hadamard state on $\aCCAR(\cV,q)$ if they are Hadamard two-point functions for $L$ and satisfy
\[
\begin{aligned}
\gi & \quad(\lambda^\pm_\sL)^*=\lambda^\pm_\sL \mbox{ \ and \ } \lambda^\pm_\sL: \ \Ran \gamma^* |_{\Gamma_{\rm c}(M;V_{[0]})}\to\Ran \gamma|_{\Gamma_{\rm c}'(M;V_{[0]})},\\
\pos & \quad \lambda^\pm_\sL  \geq 0 \mbox{ \ on \ } \Ker \gamma^* |_{\Gamma_{\rm c}(M;V_{[0]})}.
\end{aligned}
\]
\end{lemma}
\proof We have to show that $(\cdot|\lambda^\pm_{\sL}\cdot)$ induces a well-defined sesquilinear form on $\cV$. This is quite similar to the proof of Prop. \ref{eq:checkq}: it suffices to check that $(u|\lambda^\pm_{\sL} v)_{V}=0$ if $u\in\Ker \gamma^*|_{{\Gc}}$ and $v=\gamma^*f+Lh$. We have indeed
\[
(u|\lambda^\pm_{\sL} v)_{V}=(u|\lambda^\pm_{\sL} (\gamma^*f+Lh))_{V}=(\gamma^*u|\lambda^\pm_\sL f)_{V}=0,
\]
where we have used $\gi$.\qed 

\begin{remark}If the operators $\lambda^\pm_\sL$ satisfy the equations of motion and canonical (anti)-commutation relations merely `modulo gauge' in the sense that $iii)$ and $v)$ in \eqref{eq:def2p} are replaced by the weaker conditions
\beq\label{eq:moddafs}
\begin{aligned}
iii')&\quad \lambda^{\pm}_\sL L=0 \ \mbox{modulo\ operators\ that\ map\ to\ } \Ran\gamma|_\Gamma, \\[1mm]
v')&\quad \lambda^{+}_\sL \mp(-1)^{\rm gh}\lambda^{-}_\sL = \i G_\sL  \ \mbox{modulo\ operators\ that\ map\ to\ } \Ran\gamma|_\Gamma, 
\end{aligned}
\eeq
then $\lambda^\pm_\sL$ can  still  be used to define two-point functions of a Hadamard state on $\aCCAR(\cV,q)$. However, the main benefit of a two-point function is that it can be used for the deformation quantization of an extended algebra containing nonlinear local functionals \cite{DutschFredenhagenDeformation}. The weaker conditions \eqref{eq:moddafs} do not seem to ensure that this deformation quantization fulfills the basic commutator axiom on non-linear functionals. 
\end{remark}

To sum up, to construct Hadamard states in the BRST formalism one is left in practice with finding two-point functions for $L$ (i.e. $\lambda^\pm_\sL:\Gamma_{\rm c}(M; V)\to \Gamma_{\rm c}'(M; V)$ satisfying $i)$--$v)$ in (\ref{eq:def2p})) that satisfy additionally the \emph{gauge-invariance} condition $\gi$, \emph{positivity} $\pos$ and the \emph{Hadamard condition} $\musc$.    

The relation between the BRST formalism and the subsidiary condition framework, explained in Subsect. \ref{ss:relation}, can be extended to states. This is expressed in more precise terms in the following easy proposition, which formalises an argument given in \cite{hollands} for the Yang-Mills equation linearized around a flat connection on a trivial bundle (the same argument is also used in \cite{FS} for the Maxwell equation). We state only a version for bosonic theories as there are no good examples of fermionic theories satisfying the assumptions of the subsidiary condition framework with $T=K$.

\begin{proposition}\label{prop:hadbrs}Suppose $P\in\Diff(M;V_1)$ and $K\in\Diff(M;V_0,V_1)$ satisfy Hypothesis \ref{as:subsidiary} with $K=T$, in particular $D\defeq P+KK^*$ and $R\defeq K^*K$ are Green hyperbolic.

Let $V$ and $L,\gamma\in\Diff(M;V)$ be defined as in Subsect. \ref{ss:relation} with $\alpha=1$.

Suppose $\lambda_\sD^\pm$, $\lambda_\sR^\pm$ are bosonic Hadamard two-point functions for $D$, $R$, s.t.
\beq\label{eq:stsub}
\begin{aligned}
& K\lambda^\pm_\sR = \lambda^\pm_\sD K,\\
& \lambda^\pm_\sD \geq 0 \mbox{ on } \Ker K^*|_{\Gc},
\end{aligned}
\eeq
Then
\beq\label{eq:LL}
\lambda_\sL^+\defeq \begin{pmatrix}
 \lambda_\sD^+  & K \lambda_\sR^+ &  0 & 0 \\
   \lambda_\sR^+ K^* & 0 & 0 & 0 \\
 0 & 0 & 0 & \lambda_\sR^+ \\
 0 & 0 & \lambda_\sR^+ & 0
\end{pmatrix}, 
\quad 
\lambda_\sL^-\defeq \begin{pmatrix}
 \lambda_\sD^-  & K \lambda_\sR^- &  0 & 0 \\
 \lambda_\sR^- K^*   & 0 & 0 & 0 \\
 0 & 0 & 0 & -\lambda_\sR^- \\
 0 & 0 & -\lambda_\sR^- & 0
\end{pmatrix}
\eeq
are bosonic two-point functions for $L$ and two-point functions of a Hadamard state on $\aCCR(\cV,q)$.
\end{proposition}

Simple computations show that $\lambda^\pm_\sL$ satisfy conditions $i)$ to $v)$ in (\ref{eq:def2p}) and $\gi$, $\musc$ and  $\pos$ indeed. 
 
Examples of Hadamard two-point functions $\lambda^\pm_\sR$, $\lambda^\pm_\sD$ satisfying (\ref{eq:stsub}) are constructed under various topological assumptions in \cite{GW2} for the Yang-Mills equation linearized around a space-compact solution, in \cite{hollands} for the Yang-Mills equation linearized around a flat connection and in \cite{FP,FS,DS} for the Maxwell equation.

Combined with Prop.~\ref{prop:hadbrs}, this yields a construction of Hadamard states for the Maxwell and Yang-Mills theory in the BRST framework.


\subsection{Quantization}\label{ss:quantization}

In what follows we briefly discuss algebraic quantization in  the BRST formalism in order to make the connection with the terminology used in the literature.

Suppose that we have Hadamard two-point functions $\lambda^\pm_\sL$ that satisfy conditions $\gi$ and $\pos$ from Lemma \ref{lem:obvious}. These define uniquely a state on $\aCCAR(\cV,q)$, and one can use the GNS construction in the standard way to get field operators on a Hilbert space $\cH$.

In practice, however, it is more convenient to work with the `unphysical' phase space $(\cV_\sL,q_\sL)$ and its Cauchy surface version $(\cV_{\sL\Sig},q_{\sL\Sig})$, which simply consists of test sections (instead of being a quotient of spaces like $\cV_\Sig$). Thus, one views $\lambda^\pm_\sL$ as the two-point function of a \emph{pseudo-state} (i.e., a non-necessarily positive unital functional) on a bigger algebra $\aGH(\cV_\sL,q_{\sL})$ or $\aGH(\cV_{\sL\Sig},q_{\sL\Sig})$. This $*$-algebra is defined as $\aCCR$ and $\aCAR$, except that it uses the grading $\gh$ to distinguish between bosonic and fermionic degrees of freedom (as in $v)$ of (\ref{eq:def2p})).

An appropriate generalisation of the GNS construction (see for instance \cite{hofmann}) produces operators on a topological vector space $\cK$ and an indefinite inner product $(\cdot|\cdot)$ on $\cK$. The BRST operator $\gamma$ is promoted to an operator $\hat\gamma$ on $\cK$, and the `physical Hilbert space' is defined to be $(\cH, (\cdot|\cdot) )$ where
\[
\cH\defeq \frac{\Ker \hat\gamma}{\Ran \hat\gamma}.
\]
Often in the literature, one states the following conditions that ensure that $\cH$ is a pre-Hilbert space and the physical observables are faithfully represented (see for instance \cite{DuetschFredenhagen}):
\[
\begin{aligned}
{\rm \textit{i})}& \ \ (f|f)\geq 0 \ \ \forall f \in\Ker \hat\gamma,\\
{\rm \textit{ii})} & \ \ (f|f)=0, \ \ f\in\Ker \hat\gamma \ \Leftrightarrow \ f\in \Ran \hat\gamma.
\end{aligned}
\]
Condition ${\rm \textit{i})}$ is equivalent to our positivity condition $\pos$.

The implication $\Leftarrow$ in condition ${\rm \textit{ii})}$ is implied by the gauge invariance condition $\gi$.

The implication $\Rightarrow$, however, is more delicate and requires that 
\beq\label{eq:lamppm}
(\cdot| (\lambda^+_\sL  + \lambda^-_\sL)\cdot)
\eeq 
is non-degenerate on $\cV$ (resp. $(\cdot| (\lambda^+_\sL  - \lambda^-_\sL)\cdot)$ in the fermionic case). This follows by construction and from the fact that non-degeneracy of (\ref{eq:lamppm}) is equivalent to the faithfulness of the corresponding pseudo-state\footnote{This is more easily seen in the real setting, since $\lambda^+_\sL  + \lambda^-_\sL$ is (proportional to) the complexification of the real covariance.}. It appears that non-degeneracy of (\ref{eq:lamppm}) is an issue when the physical phase space $(\cV,q)$ is degenerate. Indeed, we have seen in Remark \ref{rem:rem} that typically, degeneracy of $q_\Sig$ on $\cV_\Sig$ entails that any hermitian form such as (\ref{eq:lamppm}) is degenerate. 

\section{Examples and applications}\label{s:examples}

\subsection{Maxwell equation}\label{ss:max}

The quantization of the Maxwell equation in the subsidiary condition framework was considered in many works, its relation to the BRST framework was also discussed in \cite{hollands,FS}.  

In short, one shows that Hypothesis \ref{as:subsidiary} is satisfied by
\beq\label{eq:maxwell}
P=\delta d \in \Diff^2(M;\Lambda^1), \quad K=d \in\Diff^1(M;\Lambda^0,\Lambda^1),
\eeq
and $T=K$. Above, $\Lambda^i$ is the bundle of $i$-forms on $M$, $d$ is the differential and $\delta$ the codifferential.

The purpose of this section is to make the connection between the criterion for non-degeneracy from Subsect.~\ref{ss:nond} and known results about $L^2$-cohomology of the differential $d_\Sig$ on $\Sigma$.

We will use the 
notation  $H^i_{*}(d_\Sig)$, $H_{i,*}(\delta_\Sig)$ introduced in \ref{sss:coh} for the respective (co)homologies.

Recall that on $i$-forms $\Lambda^i(\Sigma)$, using the Hodge operator $\star:\Lambda^i(\Sigma)\to\Lambda^{d-i}(\Sigma)$ one defines a scalar product
\beq\label{eq:schodge}
(u|v)\defeq \int_\Sigma u \wedge \star v\, d{\rm Vol}_{g}, \quad u,v\in\Lambda^i(\Sigma).
\eeq
The codifferential $\delta_\Sig$ is then the formal adjoint of $d_\Sig$ for this scalar product. We denote $\Delta\defeq \delta_\Sig d_\Sig\in\Diff^2(\Sigma;\Lambda^0)$ the Hodge Laplacian on $0$-forms.

The embedding of $\Ker d_\Sig|_{\Gamma(\Sigma; \Lambda^i)}$ into 
\[
\{ u\in \Gc' :  (u|v)=0 \ \forall\,v\in\Ran\delta_\Sig|_{\Gc(\Sigma;\Lambda^{i+1})}\}
\]
 induces a map
\beq\label{eq:poincare}
H^i(d_\Sig)\stackrel{\tilde{\jmath}}{\longrightarrow}\big( H_{i,\c}(\delta_\Sig)\big)^*.
\eeq
In this terminology, Poincar\'e duality says that $\tilde{\jmath}$ is injective\footnote{This follows from the usual formulation of Poincar\'e duality and basic properties of the Hodge $\star$ operator.}.

\begin{proposition}\label{prop:ndgmax}In the case of the Maxwell equation (\ref{eq:maxwell}),
 $q_\Sig$ is non-degenerate on $\cV_\Sig$ iff the canonical map 
 \[
 H^1_{\rm c}(d_\Sig)\stackrel{\tilde{\imath}}{\longrightarrow} H^1(d_\Sig)
 \]
  is injective.
\end{proposition}
\proof First, we will need a result from \cite{GW2} which states that for a convenient choice of Cauchy data, the operator $K_\Sig$ defined in (\ref{eq:defKsig}) can be expressed as
\[
K_{\Sig}= \left(\begin{array}{cc}
0&\i\\
{d_\Sig}&0\\
0&0\\
0&0
\end{array}\right), \quad K_{\Sig}^{\dag}=\left(
\begin{array}{cccc}
0&0&\i&0\\
0&0&0&{\delta_\Sig}
\end{array}\right).
\]
Using the results of Subsect.~\ref{ss:relation}, we obtain that in the BRST framework
\[
\begin{aligned}
(\Ker \gamma_\Sig|_{\Gc})\gho &= \Gamma_\c(\Sigma;\Lambda^0)\oplus\Gamma_\c(\Sigma;\Lambda^1)\oplus \{0\} \oplus \Ker \delta_\Sig|_{\Gc},\\
(\Ran \gamma_\Sig^*|_{\Gc})\gho &= \{0\}\oplus\{0\} \oplus \Gamma_\c(\Sigma;\Lambda^0) \oplus \Ran d_\Sig|_{\Gc}
\end{aligned}
\]
and analogous identities hold for $\Gamma_{\rm c}$ and $L^2$. From this point on it is straightforward to check that the injectivity of canonical maps for $d_\Sig$-cohomology (including (\ref{eq:poincare})) entail analogous properties of $\gamma_\Sig$-homology. Therefore, the claim follows from Thm. \ref{thm:criterion}.\qeds

We thus see that Thm.~\ref{thm:criterion} reduces in this case to the result from \cite[Prop. 3.5]{SDH}.
 
\begin{remark} Several references discuss an injectivity condition between  $H^1_{\rm c}(d_\Sig)$ and the so-called \emph{reduced $L^2$ cohomology} $H^1_{L^2}(d_\Sig)$ \cite{carron0,carron,mazzeo,litam}. One may ask what additional conditions ensure that smooth representatives of $H^1_{L^2}(d_\Sig)$ are injectively embedded in $H^1(d_\Sig)$. 
Unfortunately, one obtains this way sufficient conditions for non-degeneracy that cover only partially the examples discussed in \cite{SDH}. It is also possible to define in general a reduced $L^2$-cohomology for $\gamma_\Sig^*$ (not only in the Maxwell case), its study is however more difficult due to the fact that its equivalence classes do not necessarily have smooth representatives.\end{remark} 
 


\subsection{Linearized Yang-Mills equation}\label{ss:YM}

Let us briefly discuss the case of the Yang-Mills equation linearized around a generic smooth solution $\wbar{A}$, which is a connection on a principal $G$ bundle $B$ over $M$. We only assume that $(M, g)$ is globally hyperbolic. For details on the geometric constructions, we refer to \cite{KN, MM}. Our purpose will be to give examples for degeneracy of $q$.

The space of connections on a principal bundle is an affine space, with associated linear space $\Gamma(M; E^1)$, where $E^i \defeq (B \times_\ad \fg) \otimes \Lambda^i$. One defines the \emph{exterior product}
\[
 (a \otimes \omega) \wedge (b \otimes \nu) \defeq  [a, b] \otimes (\omega \wedge \nu) \quad a,b \in \Gamma(M; E^0), \ \omega, \nu \in \Gamma(M; \Lambda ).
\]
The connection $\wbar{A}$ induces a covariant derivative $\wbar{\nabla}$ on $\Gamma(M;E^0)$. We define the \emph{covariant differential} $\wbar{\ud}: \Gamma(M;E^k) \to \Gamma(M;E^{k+1})$ by
\[
 \wbar \ud (a \otimes \omega)  = \wbar{\ud} a \otimes \omega + a \otimes \ud \omega, \quad (\wbar \ud a)(X)  \defeq \wbar \nabla_X a. 
\]
Note that this is in general not a differential, but
\begin{equation*}
 \wbar \ud \circ \wbar \ud = \wbar F \wedge\,,
\end{equation*}
where $\wbar F$ is the curvature of $\wbar{A}$.

We may also define the Hodge operator $\star: \Gamma(M; E^k) \to \Gamma(M; E^{n-k})$ by
\[
 \star(a \otimes \omega) \defeq a \otimes ( \star \omega).
\]
There is a natural pairing $\Gamma(M; E^k) \times \Gamma(M;E^l) \to C^\infty(M)$ defined by
\[
 \scalP{a \otimes \omega}{b \otimes \nu} \defeq \scalP{a}{b}_{\killing} \scalP{\omega}{\nu}_g,
\]
where $\scalP{\cdot}{\cdot}_{\killing}$ is the pairing $\Gamma(M; E^0) \times \Gamma(M;E^0) \to C^\infty(M)$ induced by the Killing form on the fibers, and $\scalP{\cdot}{\cdot}_{g}$ is the pairing of forms induced by the metric $g$. Composition with integration yields a scalar product
\[
 (a \otimes \omega | b \otimes \nu ) \defeq \int \scalP{a \otimes \omega}{b \otimes \nu} d{\rm Vol}_{g},
\]
which is well-defined on $\Gamma_\c(M;E^k)$. As usual, the exterior product and $\wbar \ud$ have adjoints \wrt the scalar product, namely the interior product and $\wbar \delta$, given by
\[
 \wbar \delta  \defeq (-1)^{n(k+1)+1} \star \circ \wbar \ud \circ \star, \quad (a \otimes \omega) \Int (b \otimes \nu)  \defeq [b, a] \otimes \omega \Int \nu.
\]
That $\wbar \delta$ is indeed the adjoint of $\wbar \ud$ follows from the fact that the covariant derivative $\wbar \nabla$ is metric \wrt the pairing $\scalP{\cdot}{\cdot}_\killing$.

In the language of the subsidiary condition framework, the operators $P$ and $K=T$ are given by
\[
 P  = \wbar \delta \wbar \ud + \wbar F \tnI, \quad K = \wbar \ud,
\]
where $\tnI$ is defined by $\wbar F \tnI A \defeq A \Int \wbar F$. Appropriate Cauchy data maps for sections of $E^1(M)$ are generalizations of those given in \cite{furlani}, c.f.\ also \cite{GW2} for the case of trivial bundles over static spacetimes:
\begin{align*}
 \rho_0 & \defeq \iota^\pback, \\
 \rho_{\wbar \ud} & \defeq \i^{-1} (-1)^{p(n-p-1)+n-1} \star_\Sig \iota^\pback \star \wbar \ud, \\ 
 \rho_{\wbar \delta} & \defeq \iota^\pback \wbar \delta, \\
 \rho_{n} & \defeq \i^{-1} (-1)^{p(n-p-1)+n-1} \star_\Sig \iota^\pback \star.
\end{align*}
Here $\iota^\pback$ is the pullback along the embedding $\iota: \Sigma \to M$, and for later convenience, we have stated the maps as acting on sections of $E^p$. Obviously, the tuple $(\rho_n, \rho_0, \rho_{\wbar \delta}, \rho_{\wbar \ud})$ maps to $\Gamma_\c(\Sigma; E^0(\Sigma) \oplus E^1(\Sigma) \oplus E^0(\Sigma) \oplus E^1(\Sigma))$. Furthermore, $(\rho_0, \rho_{\wbar \ud})$ is a Cauchy data map for sections of $E^0(M)$ and the wave operator $K^* K$. The representation of the operator $K$ on the Cauchy data is then given by
\[
K_{\Sig}= \left(\begin{array}{cc}
0&\i\\
{\bds}&0\\
0&0\\
\i^{-1}a&0
\end{array}\right), \quad K_{\Sig}^{\dag}=\left(
\begin{array}{cccc}
0&0&\i&0\\
0&\i\,a^{*}&0&{\bdeltas}
\end{array}\right),
\]
where $a \defeq \rho_n \wbar F \wedge$ and $\bds$ (resp. $\bdeltas$) is the differential (codifferential) associated to the connection $\wbar A_\Sig$ induced by $\wbar A$. In particular, we have,
\begin{equation*}
 (\Ran \gamma^*_\Sig|_{\Gamma_{(\c)}})|_{[0]} = \left\{ (f, \wbar d_\Sig g, 0, \i^{-1} a g) : \ f, g \in \Gamma_{(\c)}(\Sigma; E^0) \right\}
\end{equation*}
Hence, injectivity of $\imath$, c.f.\ \eqref{eq:coh1}, is violated iff there is some $g \in \Gamma(\Sigma; E^0)$ such that $(\wbar d_\Sig g, \i^{-1} a g) \in \Gamma_\c(\Sigma; E^1)$ and there is no $g' \in \Gamma_\c(\Sigma; E^0)$ such that $(\wbar d_\Sig g, \i^{-1} a g) = (\wbar d_\Sig g', \i^{-1} a g')$. 

For a non-Abelian gauge group, it is straightforward to devise examples of violation of injectivity even on a topologically trivial spacetime. Take $M$ as Minkowski space-time and a global trivialization, so that the connection may be expressed as a $\fg$-valued one-form $\wbar A$. Take $\Sigma$ as a fixed time surface. Now choose non-trivial initial data on $\Sigma$, with support contained in a compact region $X$, with $\wbar A_0 = 0$ and $\partial_t \wbar A_\mu = 0$ (these satisfy the constraint equations, c.f.\ \cite{Segal}). These initial data determine a global smooth solution \cite{CS,Segal}, with $\rho_0 \wbar F = \wbar F_\Sig$ non-trivial with support in $X$ and $\rho_n \wbar F=0$ (which means $a=0$). Take a section $g \in \Gamma(\Sigma; E^0)$ which is covariantly constant on $\Sigma \setminus X$, i.e. $\wbar d_\Sig g|_{\Sigma\setminus X}=0$. Obviously, $(\wbar d_\Sig g, \i^{-1} a g)=(\wbar d_\Sig g, 0)$ is compactly supported. But unless $g|_{\Sigma \setminus X}$ can be extended to a covariantly constant section on $\Sigma$, we can not write it as $(\wbar d_\Sig g', 0)$ for some compactly supported $g'$. Indeed for a suitably chosen $g$, such an extension is not possible, due to the curvature of $\wbar F_\Sig$ (there is a basis $\{ g_i \}$ of covariantly constant sections on $\Sigma \setminus X$, but not on $\Sigma$, as the requirements $\wbar \ud_\Sig g_i = 0$ and $\wbar \ud_\Sig \wbar \ud_\Sig g_i = \wbar F_\Sig \wedge g_i$ are in general incompatible). 

Heuristically, one can say that in electrodynamics, a failure of injectivity can occur because charges can be hidden in a `hole'. In non-Abelian gauge theory, charges can also be hidden in regions where the gauge field is nontrivial.

\subsection{Rarita-Schwinger equation}\label{ss:RS}

We now discuss the Rarita-Schwinger equation, c.f.\ \cite{Nilles84, WeinbergIII}.
In \cite{HS} it is shown how it fits into the subsidiary condition framework with $T\neq K$. The purpose of this section is to compare the method from \cite{HS} with more conventional BRST-based approaches.

The original massless Rarita-Schwinger equation is
\[
 (P_{\RS} \Psi)^\mu \defeq  \gamma^{\mu \nu \lambda} \nabla_\nu \psi_\lambda = 0,
\]
where $\gamma^{\mu \nu \lambda}$ stands for the completely antisymmetrized product of $\gamma^\mu$, $\gamma^\nu$ and $\gamma^\lambda$. Here $\psi_\nu$ is a section of $V_1 = V_0 \otimes T^*M$, where $V_0 \defeq \DM$ is the standard Majorana bundle corresponding to a spin structure ${\rm\textit{SM}}$ over $M$, c.f.~\cite{HS} for details. Note that with our convention on the Lorentzian signature gamma matrices are anti-hermitian and so $P_{\RS}$ is formally self-adjoint. This bundle is equipped with a natural anti-symmetric bilinear form $(\cdot|\cdot)_{V_1}$ induced by the canonical hermitian structures of the Dirac and the cotangent bundle. As in \cite{HS}, we assume $(M,g)$ is a Ricci-flat spacetime of dimension $n$, $n\geq3$. In this case there is a gauge symmetry given by
\[
 (K_{\RS} \phi)_\mu = \nabla_\mu \phi, \quad \phi\in\Gc(M;\DM).
\]
Instead of working with the field $\psi$ and the Rarita-Schwinger equation, it was proposed in \cite{ET} to consider the field
\[
 \psi_\mu = (F^{-1} \Psi)_\mu
\]
where $F$ is the formally self-adjoint operator given by
\begin{align*}
 (F \psi)_\mu & \defeq \psi_\mu - \tfrac{1}{n-2} \gamma_\mu \gamma^\nu \psi_\nu, \\
 (F^{-1} \psi)_\mu & = \psi_\mu - \tfrac{1}{2} \gamma_\mu \gamma^\nu \psi_\nu.
\end{align*}
The natural equations of motion for $\phi$ are given by the operator
\[
 P \defeq F^* \circ P_{\RS} \circ F,
\]
i.e.,
\beq\label{eq:defP}
 (P \psi)_\mu =  (\nabslash \psi_\mu + \tfrac{1}{n-2} \gamma_\mu \nabslash \gamma^\nu \psi_\nu)
\eeq
Furthermore, one can define the gauge transformation operator as
\[
 K \defeq F^{-1} \circ K_{\RS},
\]
so that $P \circ K = 0$ is ensured. Concretely,
\begin{align}
\label{eq:RSPK}
(K \phi)_\mu & = \nabla_\mu \phi - \tfrac{1}{2} \gamma_\mu \nabslash \phi, \\
K^* \psi & = - \nabla^\mu \psi_\mu + \tfrac{1}{2} \nabslash \gamma^\mu \psi_\mu. \nonumber
\end{align}
Furthermore, one can introduce the Clifford multiplication operator and its adjoint
\begin{align}
\label{eq:RSPK2}
(T\phi)_\mu & \defeq -\gamma_\mu \phi, \\
T^* \psi & \defeq \gamma^\mu \psi_\mu. \nonumber
\end{align}
The Dirac operator acting on sections of $V_0$ can then be written as
\begin{equation}
\label{eq:DiracOp}
 \nabslash = - \tfrac{2}{n-2} T^* K = - \tfrac{2}{n-2} K^* T.
\end{equation}

In \cite{HS}, a somewhat different equation of motion is considered, given by $P_{\HaS} \defeq  P_{\RS} \circ F$. Also the hermitian structure is modified accordingly, specifically they consider $(\cdot|\cdot)_{V_\HaS} \defeq ( F \cdot| \cdot )_{V_1}$. With this modified hermitian structure, $(P_\HaS, K, T)$ fulfill Hypothesis~\ref{as:subsidiary}, i.e., the conditions of the subsidiary condition framework. 


\begin{remark}The Rarita-Schwinger equation can be cast in a more geometric form as follows. Consider the bundles $F^i \defeq \DM \otimes \Lambda^i(M)$. One introduces a covariant derivative, induced by the spin connection, as 
\[
 \wbar d (\psi \otimes \omega) \defeq \bar \nabla_\mu \psi \otimes d x^\mu \wedge \omega + \psi \otimes d \omega,
\]
and the Clifford multiplication as 
\[
 \Gamma (\psi \otimes \omega) \defeq \Gamma_\mu \psi \otimes \ud x^\mu \wedge \omega.
\]
Furthermore, one defines the Hodge dual as
\[
 \star (\psi \otimes \omega) \defeq \psi \otimes \star \omega.
\]
Then the operator $P$ is proportional to $\star \bar d \star \Gamma$ and $K^*$ is proportional to $\star \wbar d \star$. A similar expression for the Rarita-Schwinger equation using forms can be found in \cite{AC}.
\end{remark}


Turning our attention to the BRST framework, following \cite{EK}, we can define $L$ and $\gamma$ as
\beq\label{eq:RSPK3}
L= \begin{pmatrix}
   P & T K^* T & 0 & 0 \\
   T^* K T^* & \frac{\alpha}{2} K^* T & 0 & 0 \\
   0 & 0 & 0 & K^* T K^* T \\
   0 & 0 & K^* T K^* T & 0  \end{pmatrix},
   \quad 
   \gamma=\begin{pmatrix}
 0 & 0 &  K & 0 \\
 0 & 0 & 0 & 0 \\
 0 & 0 & 0 & 0 \\
 0 & \one & 0 & 0
\end{pmatrix}.
\eeq 
That this fulfills the requirements of the BRST framework, i.e. Hypothesis~\ref{asBRST}, follows from a direct computationand the next Proposition.
\begin{proposition}
The operator $L$ defined in (\ref{eq:RSPK3}) is Green hyperbolic.
\end{proposition}
\proof
We have
\[
 K^* T K^* T = \tfrac{(n-2)^2}{4} \nabslash \nabslash = \tfrac{(n-2)^2}{4} \nabla^\mu \nabla_\mu,
\]
which is normally hyperbolic. Hence, it remains to consider the first two rows.
Let $G^\pm_{0/1}$ be the advanced/retarded propagators for $\nabslash$ on $V_{0/1}$ and $G^\pm_\Box$ for that of $\Box = \nabslash \nabslash$ on $V_0$. We have the relations 
\begin{align}
\label{eq:Prop_1}
 \nabslash G^\pm_\Box & = G^\pm_0, \\
\label{eq:Prop_2}
 K \circ G^\pm_\Box & = -\tfrac{1}{2} G^\pm_1 \circ T, \\
\label{eq:Prop_3}
 T^* \circ G^\pm_1 \circ T & = (n-2) G^\pm_0.
\end{align}
Above, \eqref{eq:Prop_1} is well-known, \eqref{eq:Prop_2} follows from the equality $G_\sD T = K G_\sR$ of the subsidiary condition framework, and \eqref{eq:Prop_3} follows from \eqref{eq:Prop_2} and \eqref{eq:DiracOp})

Hence, if $G_\sL^\pm$ is the operator
\begin{equation*}
 \begin{pmatrix} 
 G^\pm_1 + \beta_1 G^\pm_1 \circ T \circ \nabslash \circ T^* \circ G^\pm_1 & \beta_2 K \circ G^\pm_\Box & 0 & 0 \\
 \beta_2 G^\pm_\Box \circ K^* & 0 & 0 & 0 \\
 0 & 0 & 0 & \frac{4}{(n-2)^2} G^\pm_\Box \\
 0 & 0 & \frac{4}{(n-2)^2} G^\pm_\Box & 0 
 \end{pmatrix}
\end{equation*}
with
\[
\beta_1 = \frac{1}{n-2} - \frac{\alpha}{(n-2)^3}, \quad \beta_2=\frac{4}{\alpha} (1  -  (n-2) \beta),
\]
then $L G_\sL^\pm=G_\sL^\pm L=\one$, so by its support properties $G_{\sL}^\pm$ is in fact the retarded/advanced propagator of $L$.
\qeds

Note that the propagator simplifies considerably for $\alpha = (n-2)^2$, the analogue of the Feynman gauge. For the discussion of the phase space, we thus use this particular value. In particular, our aim is to show that the phase spaces of the BRST and the subsidiary condition framework used in \cite{HS} are isomorphic. Roughly speaking, these are given by the kernel of the formal adjoint of $K$ (modulo a quotient), using however different hermitian structures. Specifically, if $K^*$ is the formal adjoint of $K$ \wrt $(\cdot|\cdot)_{V_1}$ then the formal adjoint \wrt $(\cdot|\cdot)_{V_\HaS}$ is 
\[
K^*_\HaS = K^* \circ F,
\]
so that $\Ker K^* = F \Ker K^*_\HaS$. Hence, one would expect that the isomorphism we are looking for is given by
\beq\label{eq:jziso}
 F: (\cV_\HaS, q_\HaS) \to (\cV, q),
\eeq
where $(\cV_\HaS, q_\HaS)$ is the phase space in the subsidiary condition framework associated to $P_\HaS,K,T$, and $(\cV,q)$ the phase space in the BRST framework associated to $L,\gamma$.  

Specifically, concerning the latter, we obtain by a computation as in Subsect.~\ref{ss:relation} that
\[
\cV = \frac{\Ker K^*|_{\Gc}}{\Ran P|_{\Gc}},
\]
where in one of the steps we used that $\Ran T K^* T|_{\Gamma_c} \cap \Ker K^*|_{\Gamma_c} = \{ 0 \}$ due to the fact that no compactly supported solutions to the wave equation. Furthermore, one finds 
\[
 \bar f q g = \i (f_\sA | G_1  g_\sA)_{V_1} 
\]
for any $f, g \in \Ker \gamma^*|_{\Gc(M; V_{[0]})}$, i.e., $f = (f_\sA, f_\sB) \in \Gc(M; V_1 \oplus V_0)$ and $f_\sA \in \Ker K^*$, where $G_{1}$ is the causal propagator for $\nabslash$ on $V_{1}$. 
On the other hand,
\[
\cV_\HaS = \frac{\Ker K^*_\HaS|_{\Gc}}{\Ran P_\HaS|_{\Gc}}=F\frac{\Ker K^*|_{\Gc}}{\Ran P|_{\Gc}}
\]
and
\[
 \bar{\tilde f} q_\HaS \tilde g = \i  ( \tilde f | G_1 \tilde g)_{V_\HaS}= \i  ( F \tilde f | G_1 \tilde f)_{V_1}
\]
for $\tilde f,\tilde g\in\Ker K^*_\HaS|_{\Gc}$. To prove that $\tilde f\mapsto f=F\tilde f$ is an isomorphism of phase spaces it thus remains to check that
\[
 ( F \tilde f | G_1 F \tilde g )_{V_1} = ( F \tilde f | G_1 \tilde g)_{V_1}  \qquad \forall \tilde f, \tilde g \in \Ker K^*_\HaS.
\]
We have indeed
\[
 F = \one + \tfrac{1}{n-2} T T^*,
\]
so that considering $F \tilde f \in \Ker K^*$ and \eqref{eq:Prop_2} the equality follows.

On the side note, it is worth mentioning that by \cite[Thm. 6.1]{HS}, under certain assumptions on the geometry, the hermitian form $q_{\HaS}$ in the subsidiary condition framework is positive (on $\cV_\HaS$) and therefore $(\cV_{\HaS},q_{\HaS})$ has the interpretation of a phase space of a fermionic theory\footnote{This differs from the (massive) Rarita-Schwinger field considered as a matter field (i.e. not as a gauge theory), where problems with positivity are well-known to occur, see \cite{HM}.}. By the isomorphism \eqref{eq:jziso} the same conclusion is true for the phase space in the BRST framework.\medskip

We now turn our attention to Hadamard states. A direct computation gives:
\begin{proposition}\label{prop:hadbrs2} Consider the (modified) Rarita-Schwinger operator $P$ and the operators $K,T,L$ defined in \eqref{eq:RSPK}, \eqref{eq:RSPK2}, \eqref{eq:RSPK3}. Let $D=\nabslash$, $R=\Box$. Suppose $\lambda_\sD^\pm$, $\lambda_\sR^\pm$ are fermionic, respectively bosonic Hadamard two-point functions for $D$, $R$, satisfying
\beq\label{eq:stsubss}
\begin{aligned}
& \lambda^\pm_\sD \geq 0 \mbox{ on } \Ker K^*|_{\Gc},\\
& K\lambda^\pm_\sR = - \tfrac{1}{2} \lambda^\pm_\sD T.
\end{aligned}
\eeq
Then
\begin{align}
\label{eq:LL2}
\lambda_\sL^+ & \defeq \begin{pmatrix}
 \lambda_\sD^+  & \frac{4}{(n-2)^2} K \lambda_\sR^+ &  0 & 0 \\
 \frac{4}{(n-2)^2} \lambda_\sR^+ K^*  & 0 & 0 & 0 \\
 0 & 0 & 0 & \frac{4}{(n-2)^2} \lambda_\sR^+ \\
 0 & 0 & \frac{4}{(n-2)^2} \lambda_\sR^+ & 0
\end{pmatrix}, \\
\lambda_\sL^- & \defeq \begin{pmatrix}
 \lambda_\sD^-  & \frac{4}{(n-2)^2} K \lambda_\sR^- &  0 & 0 \\
  \frac{4}{(n-2)^2} \lambda_\sR^- K^* & 0 & 0 & 0 \\
 0 & 0 & 0 & \frac{4}{(n-2)^2} \lambda_\sR^- \\
 0 & 0 & \frac{4}{(n-2)^2} \lambda_\sR^- & 0
\end{pmatrix} \nonumber
\end{align}
are fermionic two-point functions for $L$ with $\alpha = (n-2)^2$ and two-point functions of a Hadamard state on $\aCAR(\cV,q)$. 
\end{proposition}

The existence of Hadamard two-point functions $\lambda_\sD^\pm$, $\lambda_\sR^\pm$ as above requires the use of methods that lie beyond the scope of the present paper, it is however plausible that the tools developped for the Dirac and Maxwell fields \cite{SV,morehollands,FP,FS,GW2} could be generalized to solve this interesting problem.

\subsection*{Acknowledgments} The authors would like to thank Christian G\'erard, Igor Khavkine and Kasia Rejzner for useful discussions. The work of M.W.\ was partially supported by the FMJH (French Governement Program:  ANR-10-CAMP-0151-02). The work of J.Z.\ was supported by the Austrian Science Fund (FWF) under the contract P24713. The authors gratefully acknowledge the kind hospitality of the Erwin Schr\"odinger Institute during the workshop ``Algebraic Quantum Field Theory: Its Status and Its Future''.

\end{document}